\documentclass[aps,superscriptaddress,prc,twocolumn,nofootinbib,showpacs,showkeys]{revtex4-1}
\usepackage{graphicx}
\usepackage{epstopdf}
\usepackage{subfigure}
\usepackage{epsfig}
\usepackage{amsmath,amssymb,amsfonts}
\usepackage[utf8]{inputenc}
\usepackage{color}
\usepackage[bookmarks,
                bookmarksopen = true,
                bookmarksnumbered = true,
                linktocpage,
                colorlinks = true,
                linkcolor = blue,
                urlcolor  = blue,
                citecolor = blue,
                anchorcolor = green,
                hyperindex = true,
                hyperfigures]
                {hyperref}

\begin{document}

\title{Scaling behaviors of heavy flavor meson suppression and flow in different nuclear collision systems at the LHC}

\author{Shu-Qing Li}
\email{lisq79@jnxy.edu.cn}
\affiliation{School of Physical Science and Intelligent Engineering, Jining University, Qufu, Shandong, 273155, China}
\affiliation{Institute of Particle Physics and Key Laboratory of Quark and Lepton Physics (MOE), Central China Normal University, Wuhan, Hubei, 430079, China}

\author{Wen-Jing Xing}
\affiliation{Institute of Particle Physics and Key Laboratory of Quark and Lepton Physics (MOE), Central China Normal University, Wuhan, Hubei, 430079, China}

\author{Xiang-Yu Wu}
\affiliation{Institute of Particle Physics and Key Laboratory of Quark and Lepton Physics (MOE), Central China Normal University, Wuhan, Hubei, 430079, China}

\author{Shanshan Cao}
\email{shanshan.cao@sdu.edu.cn}
\affiliation{Institute of Frontier and Interdisciplinary Science, Shandong University, Qingdao, Shandong, 266237, China}

\author{Guang-You Qin}
\email{guangyou.qin@mail.ccnu.edu.cn}
\affiliation{Institute of Particle Physics and Key Laboratory of Quark and Lepton Physics (MOE), Central China Normal University, Wuhan, Hubei, 430079, China}
\date{\today}


\begin{abstract}
We explore the system size dependence of heavy-quark-QGP interaction by studying the heavy flavor meson suppression and elliptic flow in Pb-Pb, Xe-Xe, Ar-Ar and O-O collisions at the LHC. The space-time evolution of the QGP is simulated using a (3+1)-dimensional viscous hydrodynamic model, while the heavy-quark-QGP interaction is described by an improved Langevin approach that includes both collisional and radiative energy loss inside a thermal medium. Within this framework, we provides a reasonable description of the $D$ meson suppression and flow coefficients in Pb-Pb collisions, as well as predictions for both $D$ and $B$ meson observables in other collision systems yet to be measured. We find a clear hierarchy for the heavy meson suppression with respect to the size of the colliding nuclei, while their elliptic flow coefficient relies on both the system size and the geometric anisotropy of the QGP. Sizable suppression and flow are predicted for both $D$ and $B$ mesons in O-O collisions, which serve as a crucial bridge of jet quenching between large and small collision systems. Scaling behaviors between different collision systems are shown for heavy meson suppression factor and the bulk-eccentricity-rescaled heavy meson elliptic flow as functions of the number of participant nucleons in heavy-ion collisions.
\end{abstract}

\maketitle


\section{Introduction}
\label{sec:Introduction}

Quantum Chromodynamics (QCD) predicts that at extremely high temperature and density, nuclear matter transits from the hadron state to a color-deconfined state, known as the quark-gluon plasma (QGP)~\cite{Ding:2015ona}. This has been supported by a large number of experimental evidences from relativistic heavy-ion collisions at the Relativistic Heavy-Ion Collider (RHIC) and the Large Hadron Collider (LHC)~\cite{Shuryak:2014zxa}, among which anisotropic flow and jet quenching are considered as the two most important signatures of the formation of QGP. At low transverse momentum ($p_\mathrm{T}$), hadrons emitted from the QGP exhibit strong anisotropy in their azimuthal angular distributions~\cite{Adams:2003zg,Aamodt:2010pa,ATLAS:2011ah}, which has been successfully described by relativistic hydrodynamic models~\cite{Romatschke:2017ejr, Rischke:1995ir, Heinz:2013th, Gale:2013da, Huovinen:2013wma}. The small values of the specific shear viscosity extracted from hydrodynamic calculations~\cite{Song:2010mg,Bernhard:2019bmu} clearly shows the strongly-coupled nature of this fluid-like QGP. At high $p_\mathrm{T}$, hadrons and reconstructed jets emanating from initial hard scatterings are significantly quenched after traversing the QGP medium~\cite{Wang:1991xy, Gyulassy:2003mc, Majumder:2010qh, Qin:2015srf, Blaizot:2015lma, Cao:2020wlm}. The extracted large values of the jet quenching parameter $\hat{q}$~\cite{Burke:2013yra,Cao:2021keo} indicate the quark-gluon degrees of freedom inside the dense nuclear matter.

Over the past few years, large anisotropic flows have been observed in small collision systems as well, such as deuteron-gold (d-Au) collisions at RHIC and proton-lead (p-Pb) collisions at the LHC~\cite{ALICE:2012eyl, ATLAS:2012cix, CMS:2013jlh, PHENIX:2013ktj}. Interestingly, these flow coefficients can also be successfully described by hydrodynamic calculations, implying the possible formation of mini-QGP in such small systems~\cite{Bozek:2013uha, Bzdak:2013zma, Qin:2013bha, Werner:2013ipa, Bozek:2013ska, Nagle:2013lja, Schenke:2014zha, Zhao:2020wcd, Zhao:2017rgg}. On the other hand, the other strong evidence of QGP, jet quenching, has not been observed so far. For instance, despite the large elliptic flow coefficient ($v_2$) of $D$ mesons in p-Pb collisions~\cite{CMS:2018loe}, their nuclear modification factor ($R_\mathrm{pA}$) is found to be consistent with unity~\cite{ALICE:2019fhe}. This has triggered hot debates on whether the collectivity observed in small systems originates from final-state QGP effects or from initial-state gluon saturation effects~\cite{Dusling:2012iga, Dusling:2017dqg, Mace:2018vwq, Davy:2018hsl, Zhang:2019dth, Xu:2015iha, Park:2016jap, Du:2018wsj}. One possible way of disentangling the initial-state and final-state contributions to jet observables is to scan the jet quenching effect across various sizes of nuclear collision systems~\cite{Citron:2018lsq, Huang:2019tgz}. This would bridge the gap between large and small systems and may hopefully help identify the boundary across which QGP disappears. Along this direction, several theoretical efforts have been recently devoted to explore the nuclear modification effects on high $p_\mathrm{T}$ hadrons in systems smaller than Pb-Pb collisions at the LHC energies~\cite{Zigic:2018ovr, Shi:2018izg,Huss:2020dwe,Huss:2020whe}, and how parton energy loss depends on the size of collision systems~\cite{Katz:2019qwv,Liu:2021izt}.

Among various energetic probes of medium properties, heavy quarks are of particular interest~\cite{Dong:2019byy,Dong:2019unq}. Due to their large masses that suppress their thermal production from the QGP, heavy quarks are mainly produced via initial hard scatterings and then interact with the medium with their flavors conserved, which makes them a clean probe to the evolution history of the expanding QGP. Tremendous efforts have been made towards understanding the dynamics of heavy quarks inside the QGP, including their elastic scattering~\cite{He:2011qa,Das:2015ana,Song:2015ykw}, inelastic scattering~\cite{Gossiaux:2006yu,Gossiaux:2010yx,Das:2010tj,Fochler:2013epa,Cao:2017hhk,Xu:2017obm,Ke:2018tsh,Xing:2019xae,Li:2019wri,Liu:2021dpm} and hadronization~\cite{Song:2018tpv,Plumari:2017ntm,He:2019vgs,Cho:2019lxb,Cao:2019iqs} processes. For detailed comparisons between different model implementations, one may refer to Refs.~\cite{Rapp:2018qla,Cao:2018ews,Xu:2018gux,Katz:2019fkc,Li:2020kax}.

In this work, we aim at using heavy quarks to probe QGP with different sizes. Our state-of-the-art Langevin-hydrodynamics framework is applied to calculate the nuclear modification factor ($R_\mathrm{AA}$) and the elliptic flow coefficient ($v_2$) of heavy flavor mesons across Pb-Pb, Xe-Xe, Ar-Ar and O-O collisions. {The (3+1)-dimensional viscous hydrodynamic model CLVisc~\cite{Pang:2012he, Pang:2018zzo, Wu:2018cpc, Wu:2021fjf} is adopted for simulating the realistic QGP profiles produced in these collision systems, while the improved Langevin approach~\cite{Cao:2013ita,Cao:2015hia} is used for describing both elastic and inelastic scatterings of heavy quarks through the QGP medium. Hadronization plays an important role in studying heavy flavor dynamics in heavy-ion collisions. In this work, heavy quarks exiting the QGP are converted to heavy flavor hadrons using our advanced coalescence-fragmentation model~\cite{Cao:2019iqs}, which has successfully predicted the heavy flavor hadron chemistry at RHIC and the LHC. These sophisticated models on heavy quark energy loss and hadronization are necessary for a more quantitative comparison to the experimental measurements. Within this framework, we predict both $D$ and $B$ meson $R_\mathrm{AA}$ and $v_2$ for different collision systems and collision centralities, from which we investigate the hierarchy of heavy quark energy loss and its momentum anisotropy with respect to the medium size and geometric anisotropy. In particular, the scaling behaviors of heavy flavor meson observables between different collision systems are explored.
We find that both $R_\mathrm{AA}$ and the bulk-eccentricity-rescaled elliptic flow ($v_2/\varepsilon_2$) of heavy flavor mesons scale with the number of participant nucleons ($N_\mathrm{part}$).
These findings help to disentangle the effects of the overall intensity of medium modification and its geometric asymmetry on jet quenching observables, which can be tested by future measurements.}

This paper is organized as follows. In Sec.~\ref{sec:hydrodynamics}, we present the CLVisc hydrodynamic model that we use to generate the QGP profiles produced in relativistic heavy-ion collisions. In Sec.~\ref{sec:transport}, we review our Langevin approach that describes both collisional and radiative energy loss of heavy quarks inside QGP. In Sec.~\ref{sec:results}, our numerical results on $D$ and $B$ meson $R_\mathrm{AA}$ and $v_2$ are presented for different centrality regions across Pb-Pb, Xe-Xe, Ar-Ar and O-O collision systems, from which the hierarchy and scaling behaviors of these observables with respect to the system size, medium geometry and heavy quark mass will be investigated in detail. The conclusion of this study is presented in Sec.~\ref{sec:summary}.

\section{Hydrodynamic simulation of medium profiles}
\label{sec:hydrodynamics}

In this study, the dynamical evolution of the QGP medium is provided by the (3+1)-dimensional CLVisc hydrodynamic model~\cite{Pang:2018zzo,Wu:2018cpc}. The full initial entropy density distribution $S(\tau_0,x,y,\eta_s)$ is constructed by folding the smooth entropy density $s(x,y)$ in the transverse plane and the parametrized envelope function $H(\eta_s)$ in the longitudinal direction at the initial proper time $\tau_0$,
\begin{equation}
S(\tau_0,x,y,\eta_s) = Ks(x,y)H(\eta_s) |_{\tau_0},
\end{equation}
where $K$ is a scale factor which can be adjusted from the final charged hadron spectra in the most central collisions~\cite{ALICE:2018cpu,Adam:2016ddh}.
The entropy density $s(x,y)$ is generated by the Trento initial condition~\citep{Moreland:2014oya}, in which the positions of nucleons within nucleus are first sampled using the Woods-Saxon distribution,
\begin{equation}
\rho(r,\theta) = \frac{\rho_0}{1+\exp\left[\frac{r-R(\theta)}{d}\right]}\left[1+w\frac{r^2}{R(\theta)^2}\right],
\end{equation}
in which $\rho_0$ denotes the nuclear density at the nucleus center, $d$ is the surface thickness parameter, and $R(\theta)=R_0(1+\beta_2 Y_{20}(\theta) + \beta_4 Y_{40}(\theta))$ is the nuclear radius with spherical harmonic
functions $Y_{nl}(\theta)$. Here, $\beta_2$, $\beta_4$ and $w$ parameters control the deviations from a spherical nucleus. Table~\ref{tab:Woods} lists the parameters of the Woods-Saxon distribution for different nuclei used in this work.

\begin{table}[h]
\centering
\begin{tabular}{|c|c|c|c|c|c|}
\hline
Nucleus & $R_0$ [fm] & $d$ [fm] & $\omega$ & $\beta_2$ & $\beta_4$ \\ \hline
$^{208}$Pb& 6.62 & 0.546 & 0     & 0 & 0 \\ \hline
$^{129}$Xe& 5.40 & 0.590 & 0     & 0.180 & 0\\ \hline
$^{40}$Ar& 3.53 & 0.542 & 0      & 0 & 0 \\ \hline
$^{16}$O& 2.608 & 0.513 & -0.051 & 0 & 0\\ \hline
\end{tabular}
\caption{Parameters in the Woods-Saxon distribution for different collision systems~\cite{ALICE:2018cpu,Sievert:2019zjr}.}\label{tab:Woods}
\end{table}

The local entropy density $s(x,y)$ can be then constructed from the generalized mean of the nuclear matter thickness function $T_A(x,y)$ and $T_B(x,y)$ as follows:
\begin{equation}
s(x,y) = \left(\frac{T_A^p + T_B^p}{2}\right)^{\frac{1}{p}},
\end{equation}
where the thickness functions are obtained by summing over the Gaussian smearing functions (with width $0.5$~fm) of the participant nucleons inside the two colliding nuclei ($A$ and $B$).
In this work, we choose $p=0$ that corresponds to the IP-Plasma-model-like or EKRT-model-like entropy deposition. The envelope functions $H(\eta_s)$ are chosen to describe the longitudinal profile~\cite{Pang:2018zzo},
\begin{equation}
H(\eta_s) = \exp\left[ -\frac{(|\eta_s| - \eta_0)^2}{2\sigma^2_{\eta_s}}\theta(|\eta_s| - \eta_0) \right],
\end{equation}
where we use $\eta_0 = 2.23$, $\sigma_{\eta_s} = 1.8$ for Xe and $\eta_0 = 1.7$, $\sigma_{\eta_s} = 2.0$ for other nuclei in this study. For each centrality interval of each collision system, we average over 5000 Trento events to get a smooth initial entropy distribution as our hydrodynamic input at the initial proper time $\tau_0 = 0.6$~fm.

In the framework of the CLVisc hydrodynamic model~\cite{Pang:2018zzo}, the equation of motion for the energy-momentum tensor $T^{\mu\nu}$ and the dissipative equation for the shear stress tensor  $\pi^{\mu\nu}$ are solved with the
partial chemical equilibrium equation of state s95p-pce in the Milne coordinate using the Kurganov-Tadmor (KT) algorithm:
\begin{eqnarray}
& &\partial_{\mu}T^{\mu\nu}  = 0, \\
& &\pi^{\mu\nu} = \eta_v \sigma^{\mu\nu} - \tau_{\pi}\left[\Delta^{\mu\nu}_{\alpha \beta}u^{\lambda}\partial_{\lambda}\pi^{\alpha \beta} + \frac{4}{3}\pi^{\mu\nu}\theta \right],
\end{eqnarray}
where $\sigma^{\mu\nu}$ is the symmetric shear tensor and $\theta$ is the expansion rate. We set the specific shear viscosity $\eta_v/s = 0.16 $ and the relaxation time $\tau_{\pi} = 3\eta_v/(Ts)$.
After hydrodynamic evolution, the QGP is converted to hadrons via the Cooper-Frye formula with the switching temperature set as $T_{\rm sw}=137$ MeV.
With above setups, our hydrodynamic calculation provides reasonable descriptions of the soft hadron spectra in Pb-Pb collisions and Xe-Xe collisions. The QGP profiles of Ar-Ar and O-O collisions should be viewed as predictions at this moment.

\section{Heavy quark evolution inside QGP}
\label{sec:transport}

The time evolution of heavy quarks through the QGP medium is described using the modified Langevin equation~\cite{Cao:2013ita} that simultaneously includes quasi-elastic scattering and medium-induced gluon bremsstrahlung processes of heavy quarks inside a thermal medium:
 \begin{equation}
  \label{eq:Langevin}
  \frac{d\vec{p}}{dt} = -\eta _{D}(p)\vec{p}+\vec{\xi}+\vec{f_{g}}.
 \end{equation}
In the above equation, the first two terms on the right-hand side follow the classical Langevin equation, denoting the drag force and thermal random force experienced by a heavy quark while it frequently scatters with the constituents of a thermal medium. The thermal force $\vec{\xi}$ is assumed to be independent of the heavy quark momentum.
Its strength is quantified by the correlation function of a white noise $\langle\xi^{i}(t)\xi^{j}(t^{\prime})\rangle=\kappa\delta^{ij}\delta(t-t^{\prime})$, where $\kappa$ is the momentum diffusion coefficient of heavy quarks.
It is related to the spatial diffusion coefficient $D_\mathrm{s}$ via $D_\mathrm{s}\equiv T/[M\eta_{D}(0)]=2T^{2}/\kappa$, in which the fluctuation-dissipation relation $\eta_{D}(p)=\kappa/(2TE)$ is applied.

In addition to the drag and diffusion from the multiple scattering process, the effects of medium-induced gluon radiation is introduced into Eq.~(\ref{eq:Langevin}) as a recoil force $\vec{f_{g}}=-d\vec{p}_{g}/dt$ exerted on heavy quarks while they emit gluons with momentum $\vec{p}_{g}$. The probability of gluon radiation during a time interval $(t,t+\Delta t)$ is evaluated using the average number of emitted gluons during this time interval:
 \begin{equation}
 \label{eq:gluonProb}
 P_\mathrm{rad}(t,\Delta t) = \langle N_{g}(t,\Delta t)\rangle = \Delta t\int dxdk_{\perp}^{2}\frac{dN_{g}}{dxdk_{\perp}^{2}dt}.
 \end{equation}
In the calculation, we choose a sufficiently small $\Delta t$ to guarantee $\langle N_{g}(t,\Delta t)\rangle <1$, so that this average number can be utilized as a probability. In Eq.~(\ref{eq:gluonProb}), the medium-induced gluon spectrum is adopted from the higher-twist energy loss calculation~\cite{Guo:2000nz,Majumder:2009ge,Zhang:2003wk}:
\begin{equation}
\label{eq:gluonSpectrum}
\frac{dN_{g}}{dxdk_{\perp}^{2}dt}=\frac{2\alpha_{s}P(x)k_\perp^4 \hat q}{\pi ({k_{\perp}^{2}+x^{2}M^{2}})^{4}}\sin^{2}\left(\frac{t-t_{i}}{2\tau _{f}}\right),
\end{equation}
in which $x$ is the fractional energy taken by the emitted gluon from its parent heavy quark, $k_\perp$ is the transverse momentum of the gluon, $\alpha _\mathrm{s}$ is the strong coupling which runs with $k_\perp^2$ at the leading order, $P(x)$ is the $Q\rightarrow Qg$ splitting function, and $\tau _{f}=2Ex(1-x)/(k_{\perp}^{2}+x^{2}M^{2})$ denotes the splitting time with $E$ and $M$ being the energy and mass of heavy quarks respectively. Here $\hat q$ is the gluon transport coefficient which can be related to the quark diffusion coefficient via $\hat q = 2\kappa C_{A}/C_{F}$, where $C_A$ and $C_F$ are color factors of gluon and quark respectively. Note that in our modified Langevin model, there is only one free parameter which we choose as the dimensionless quantity $D_\mathrm{s}(2\pi T)$. It is adjusted as $D_\mathrm{s}(2\pi T)=4$~\cite{Li:2020kax} to provide a reasonable description of the heavy flavor meson observables in heavy-ion collisions at the LHC.

When we sample the energy-momentum of the medium-induced gluons according to Eq.~(\ref{eq:gluonSpectrum}), a lower cut-off is implemented for the gluon energy at $\omega _{0}=x_0 E = \pi T$, below which gluon is not allowed to form. Due to the lack of the gluon absorption process in the current implementation, this cut-off helps mimic the balance between gluon emission and absorption processes around the thermal energy scale. We have verified that an approximate, though not exact, thermal equilibrium of heavy quarks can be achieved after a sufficiently long time of evolution inside a thermal medium~\cite{Cao:2013ita}.

Using this Langevin framework, we can simulate the heavy quark evolution through the QGP. The realistic QGP medium is generated by the CLVisc hydrodynamic model as described in Sec.~\ref{sec:hydrodynamics}. Meanwhile, the initial heavy quarks are sampled using the binary collision vertices from the Monte-Carlo Glauber model for their position space, and the fixed-order-next-to-leading-log (FONLL) calculation~\cite{Cacciari:2001td,Cacciari:2012ny,Cacciari:2015fta} convoluted with the CT14NLO~\cite{Dulat:2015mca} parton distribution function for their momentum space. Then heavy quarks are placed into our Langevin model for their subsequent interaction with the QGP medium, which we assume to commence at the initial time ($\tau_0 = 0.6$~fm) of the hydrodynamic evolution.

After heavy quarks travel outside the QGP boundary, i.e., the local temperature of the medium drops below $T_\mathrm{c}=160$~MeV, they are converted to heavy flavor hadrons via an advanced hybrid fragmentation-coalescence model~\cite{Cao:2019iqs}. In this model, the coalescence probability between heavy quarks and thermal light quarks are calculated according to the wavefunction overlap between the free-quark state and hadronic bound state. Both $s$ and $p$-wave hadronic states are included in our calculation, which naturally cover the majority of heavy flavor hadron states observed in the Particle Data Group~\cite{ParticleDataGroup:2018ovx}. Based on this probability, heavy quarks that do not hadronize through coalescence are fragmented into heavy flavor hadrons via Pythia~\cite{Sjostrand:2006za} simulation. The heavy flavor hadrons produced from both coalescence and fragmentation processes are then utilized for analyzing the final-state observables.

As discussed earlier, a constant $D_\mathrm{s}(2\pi T)=4$ is used in this work since our main focus is on the system size dependence of heavy quark energy loss and heavy flavor suppression and flow at the LHC. One may refer to our previous study~\cite{Li:2020kax} for a detailed analysis of the systematic uncertainties introduced by various model ingredients, such as the initial heavy quark spectrum, the starting time of heavy-quark-medium interaction, the medium profile in the pre-equilibrium state, and the temperature dependence of the heavy quark diffusion coefficient, etc. {Compared to the linear Boltzmann transport model used in our earlier work~\cite{Liu:2021izt}, the Langevin approach is expected to be applicable to quasi-particles with large masses inside a thermal medium. However, it is easier to include the non-perturbative interaction between low energy heavy quarks and the QGP in the Langevin approach than in the perturbative-based Boltzmann calculation. Note that both models use the same method to implement the radiative energy loss of heavy quarks~\cite{Li:2020kax, Liu:2021izt}.}

\section{Heavy flavor meson suppression and flow}
\label{sec:results}

In this section, we present numerical results on the nuclear modifications of $D$ and $B$ mesons, and compare them between different collision systems (Pb-Pb, Xe-Xe, Ar-Ar and O-O) at the LHC energies. The two most frequently quoted heavy flavor observables -- nuclear modification factor ($R_\mathrm{AA}$) and elliptic flow coefficient ($v_2$) -- are utilized to quantify features of heavy quark energy loss inside the QGP. In the present study, they are extracted as follows from the final-state energy-momentum information of the heavy flavor mesons:
\begin{align}
\label{eq:RAA}
&R_\mathrm{AA}(p_\mathrm{T}) = \frac{1}{ N_\mathrm{coll}}\frac{dN^\mathrm{AA}/dp_\mathrm{T}}{dN^\mathrm{pp}/dp_\mathrm{T}},\\
\label{eq:v2}
&v_2(p_\mathrm{T}) = \langle \cos(2\phi)\rangle  = \left\langle \frac{p^{2}_{x} - p^{2}_{y}}{p^{2}_{x} + p^{2}_{y}}\right\rangle,
\end{align}
where $N_\mathrm{coll}$ denotes the average number of binary collisions in a given centrality bin of a given nucleus-nucleus collision system, and $\langle \ldots \rangle$ represents the average over the final-state heavy flavor mesons generated in our simulations. Smooth hydrodynamic profiles are used in this work, in which the $x$-$z$ axes define the event plane while the $x$-$y$ axes define the transverse plane of nuclear collisions. Within this setup, the azimuthal angle $\phi$ in Eq.~(\ref{eq:v2}) is measured with respect to the $+{x}$ direction. Note that each smooth hydrodynamic profile is generated from an initial entropy distribution that has been averaged over 5000 Trento events, whose participant planes have been individually rotated to the $x$-$z$ plane of our computational frame. Therefore, such smooth hydrodynamic profile has captured key features of event-by-event fluctuations in the initial state and serve as a good approximation of direct event-by-event simulations of QGP for studying heavy flavor observables. {Implementing full event-by-event calculations can lead to stronger energy loss of heavy quarks~\cite{Cao:2014fna} and larger $v_2$~\cite{Cao:2017umt} than using the smooth profiles, though such difference is expected to be within 10\%. In this work, we use the event plane method Eq.~(\ref{eq:v2}) to evaluate the heavy meson $v_2$, following the ALICE~\cite{Acharya:2017qps} and STAR~\cite{STAR:2017kkh} collaborations. Note that the correlation method has also been applied in STAR~\cite{STAR:2017kkh} and CMS~\cite{Sirunyan:2017plt} measurements. As shown by Ref.~\cite{STAR:2017kkh}, these two methods produce similar $v_2$ for heavy mesons.}

\begin{figure}[tbp]
    \centering
    \includegraphics[clip=,width=0.25\textwidth]{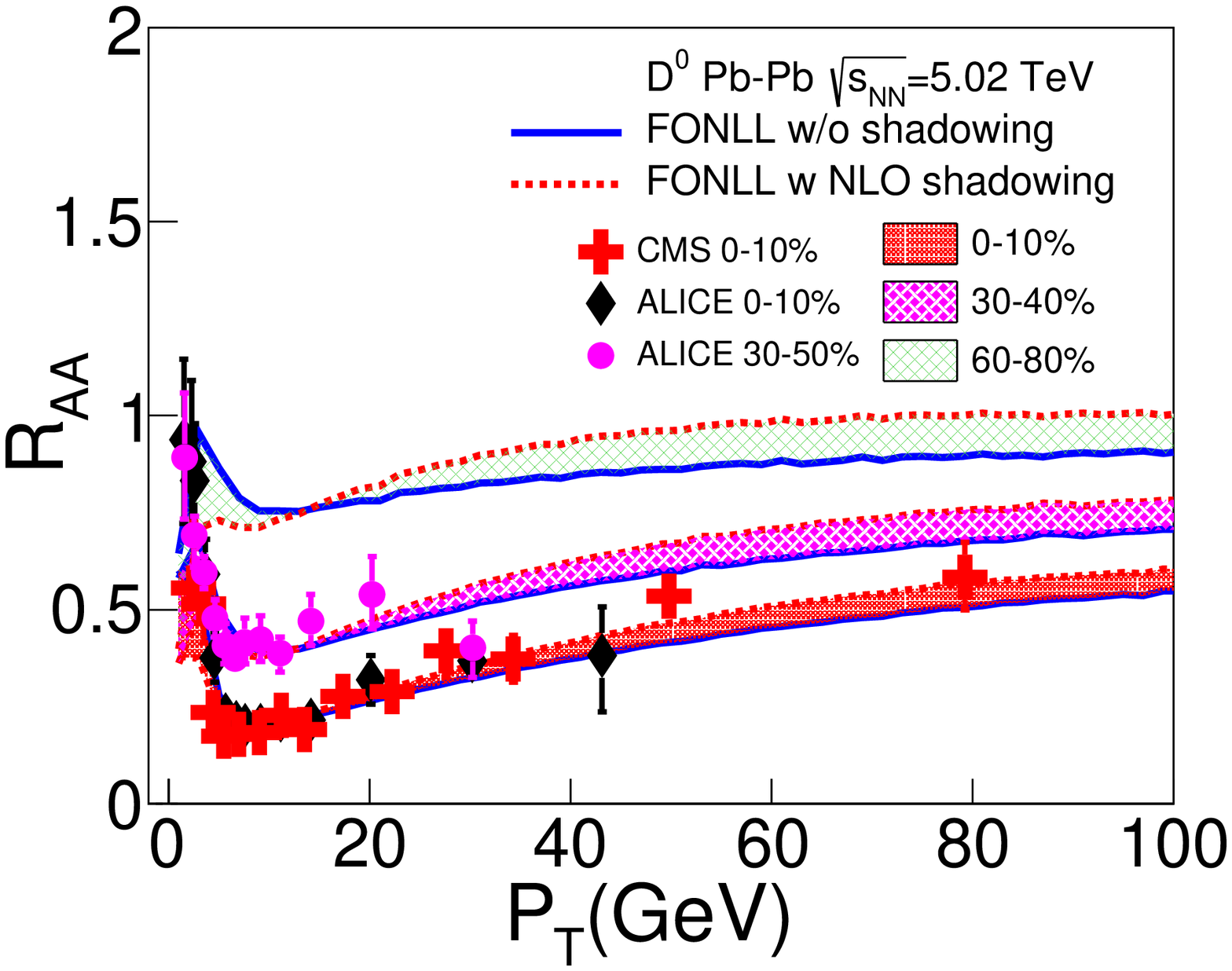}
    \hspace{-15pt}
    \includegraphics[clip=,width=0.25\textwidth]{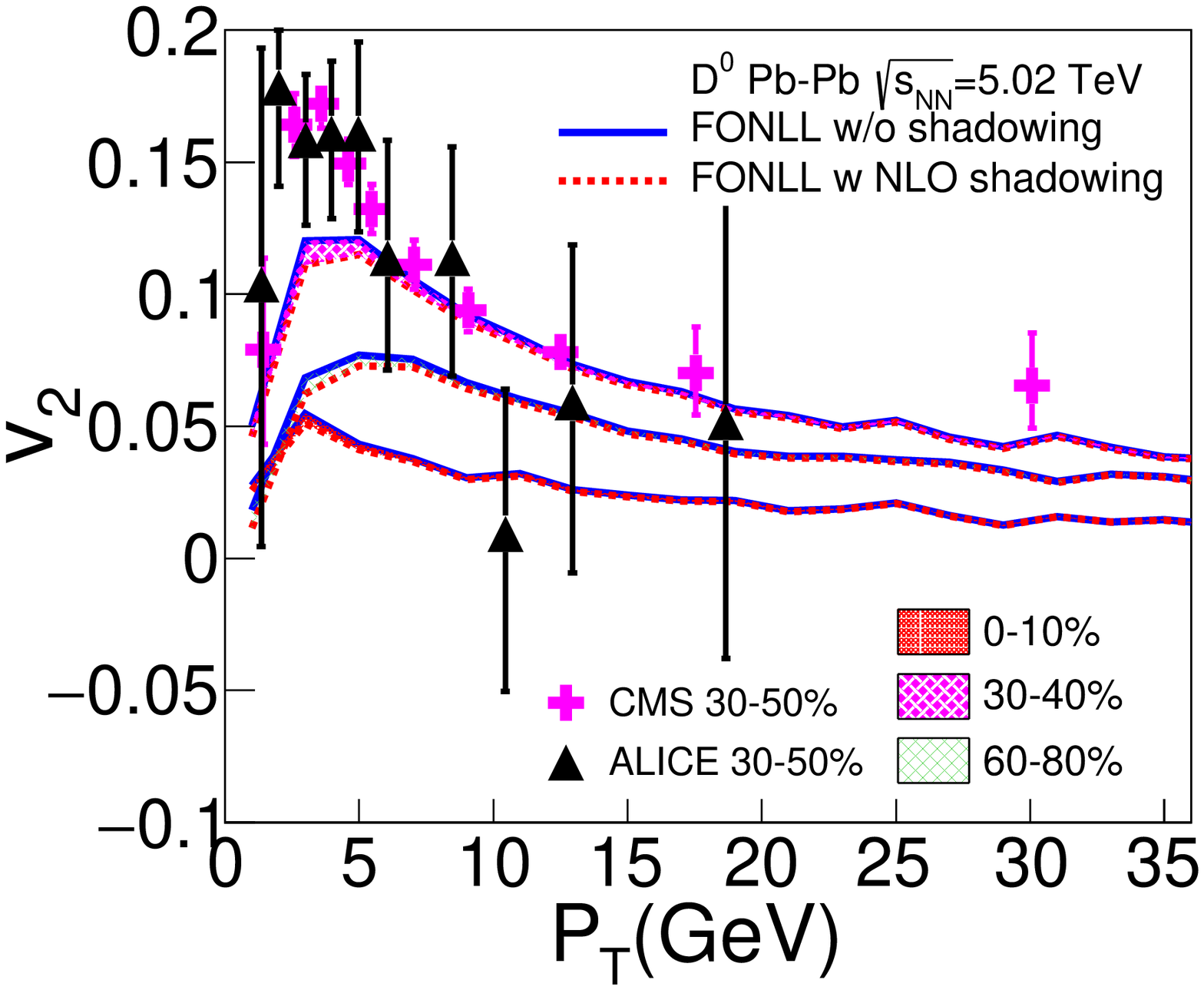}
    \includegraphics[clip=,width=0.25\textwidth]{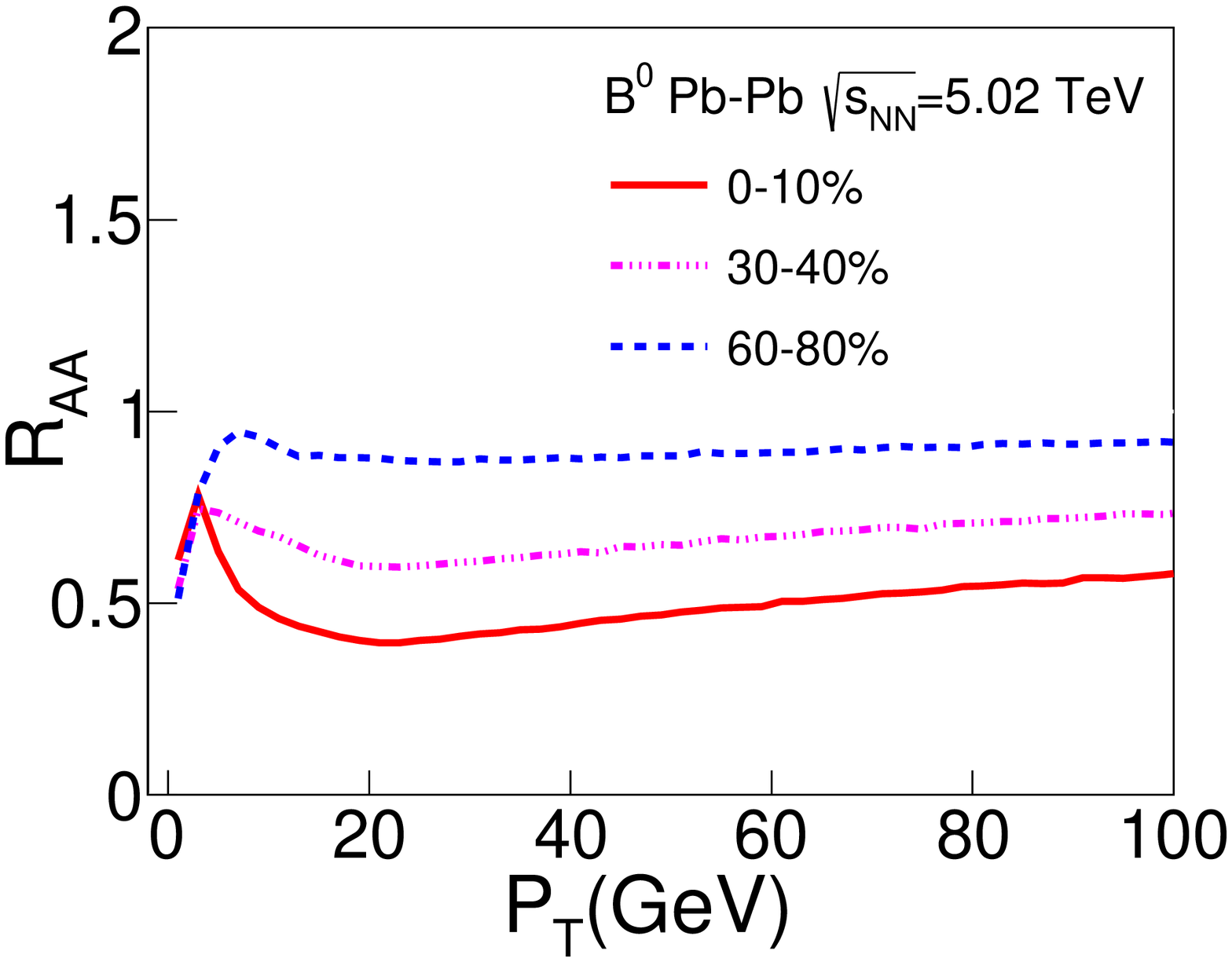}
    \hspace{-15pt}
    \includegraphics[clip=,width=0.25\textwidth]{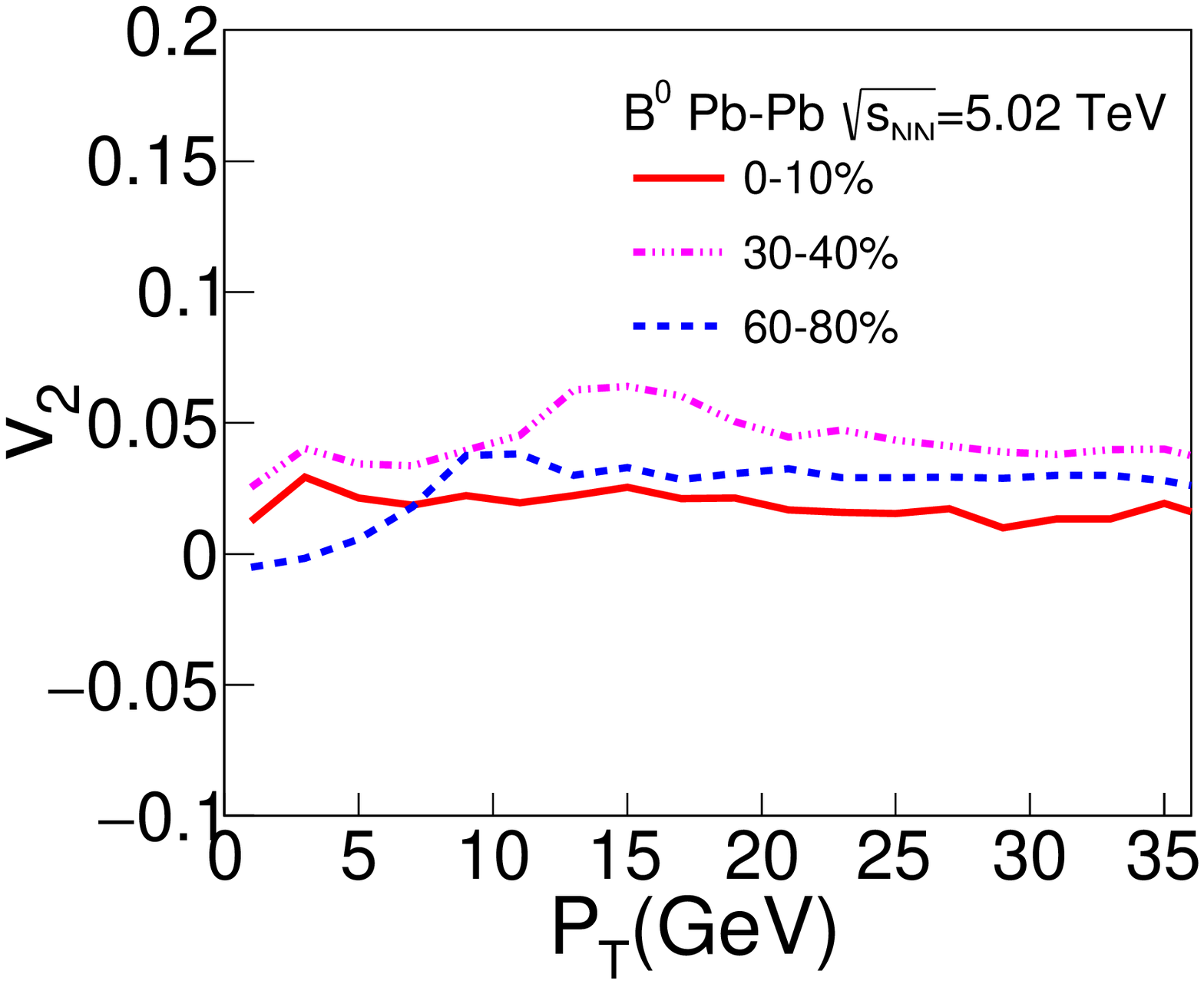}
    \caption{(Color online) Suppression (left) and elliptic flow coefficient (right) of $D$ mesons (upper) and $B$ mesons (lower) in different centrality classes of Pb-Pb collisions at $\sqrt{s_\mathrm{NN}}=5.02$~TeV.}
    \label{fig:raa-v2-pt-pbpb}
\end{figure}

With these setups, we first present in Fig.~\ref{fig:raa-v2-pt-pbpb} the $R_\mathrm{AA}$ and $v_{2}$ of $D$ and $B$ mesons in different centrality classes of Pb-Pb collisions at $\sqrt{s_\mathrm{NN}}=5.02$~TeV. In the upper panels, effects of the nuclear shadowing effects on the $D$ meson observables are shown. {The shaded bands show the difference between whether or not this nuclear shadowing effect has been included in our calculation.} After including the EPPS16~\cite{Eskola:2016oht} parametrization at the next-to-leading-order, one may observe (in the upper left panel) a suppression of the $D$ meson $R_\mathrm{AA}$ at low $p_\mathrm{T}$, while an enhancement (anti-shadowing) at high $p_\mathrm{T}$. On the other hand, this shadowing effect has little impact on the $D$ meson $v_2$, as illustrated in the upper right panel. {Note that the impact parameter averaged nuclear shadowing parametrization is used in Fig.~\ref{fig:raa-v2-pt-pbpb}. The dependence of nuclear shadowing on the impact parameter could introduce additional dependence of nuclear modification on centrality. This will be included in our future study.} Since the EPPS16 parametrization does not cover all nucleus species that we investigate in the present work, we choose to exclude this cold nuclear matter effect for the rest of our calculation in order to conduct an unbiased comparison between different collision systems. With the heavy quark diffusion coefficient set as $D(2\pi T)=4$, our results on the $D$ meson $R_\mathrm{AA}$ and $v_2$  for Pb-Pb collisions are consistent with the data from ALICE and CMS collaborations~\cite{Acharya:2018hre,Acharya:2017qps,Sirunyan:2017plt}. This helps confirm the satisfactory path-length dependence of parton energy loss embedded in our transport model.

Comparing different centrality classes in each panel of Fig.~\ref{fig:raa-v2-pt-pbpb}, one can observe a clear hierarchy in the heavy meson $R_\mathrm{AA}$, i.e., larger heavy quark energy loss in more central collisions leads to a smaller nuclear modification factor. However, this hierarchy does not hold for the heavy meson $v_2$ which depends on the competing effects between the amount of energy loss and the geometric anisotropy of the medium. The former is stronger in more central collisions, while the latter is larger in more peripheral collisions. Therefore, one usually observes a maximum for elliptic flow $v_2$ in semi-central/peripheral collisions (e.g. 30-40\%). Comparing upper panels and lower panels, one can observe that $D$ mesons have smaller $R_\mathrm{AA}$ and larger $v_2$ than $B$ mesons because charm quarks have much smaller mass than bottom quarks.

\begin{figure}[tbp]
    \centering
    \includegraphics[clip=,width=0.25\textwidth]{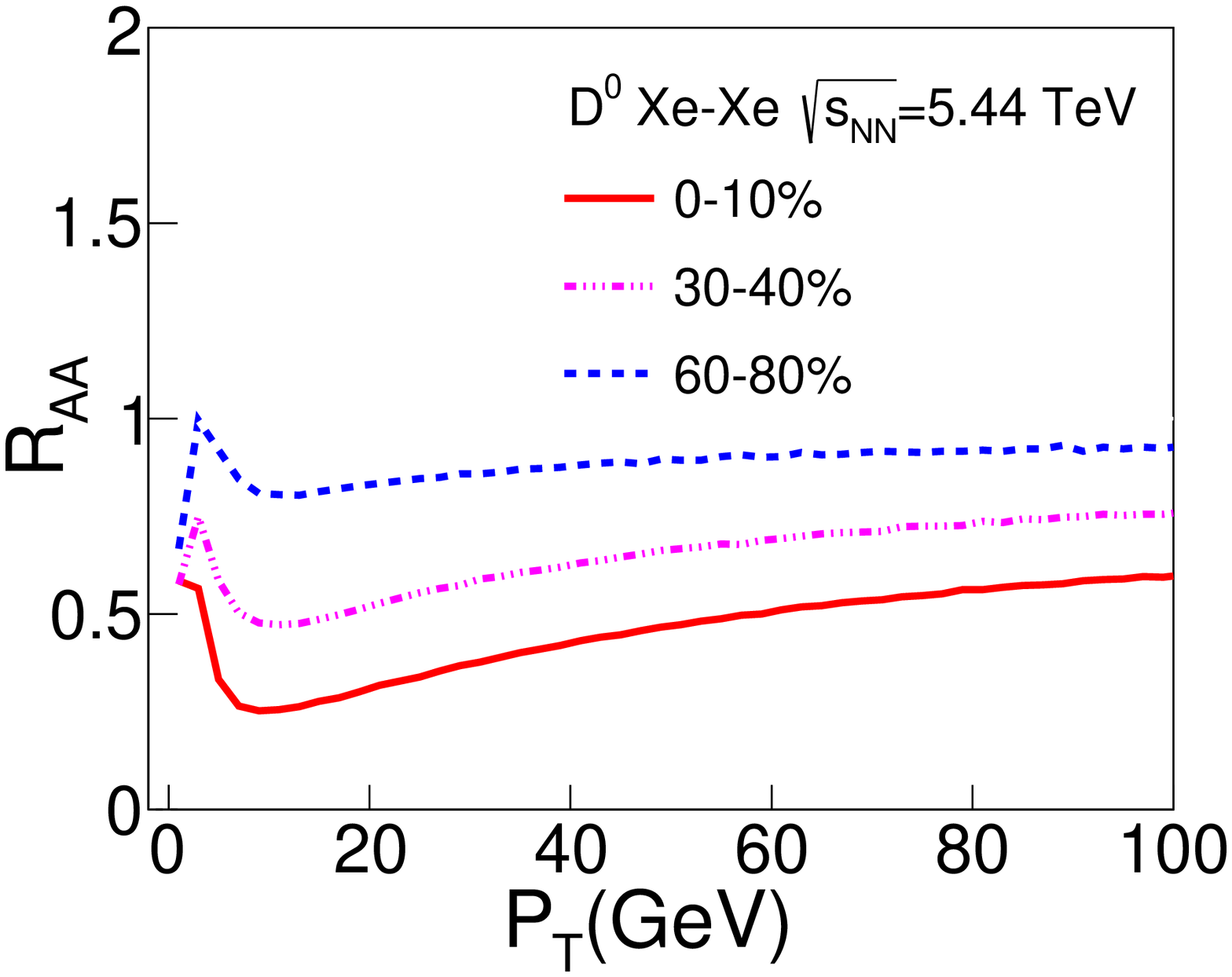}
    \hspace{-15pt}
    \includegraphics[clip=,width=0.25\textwidth]{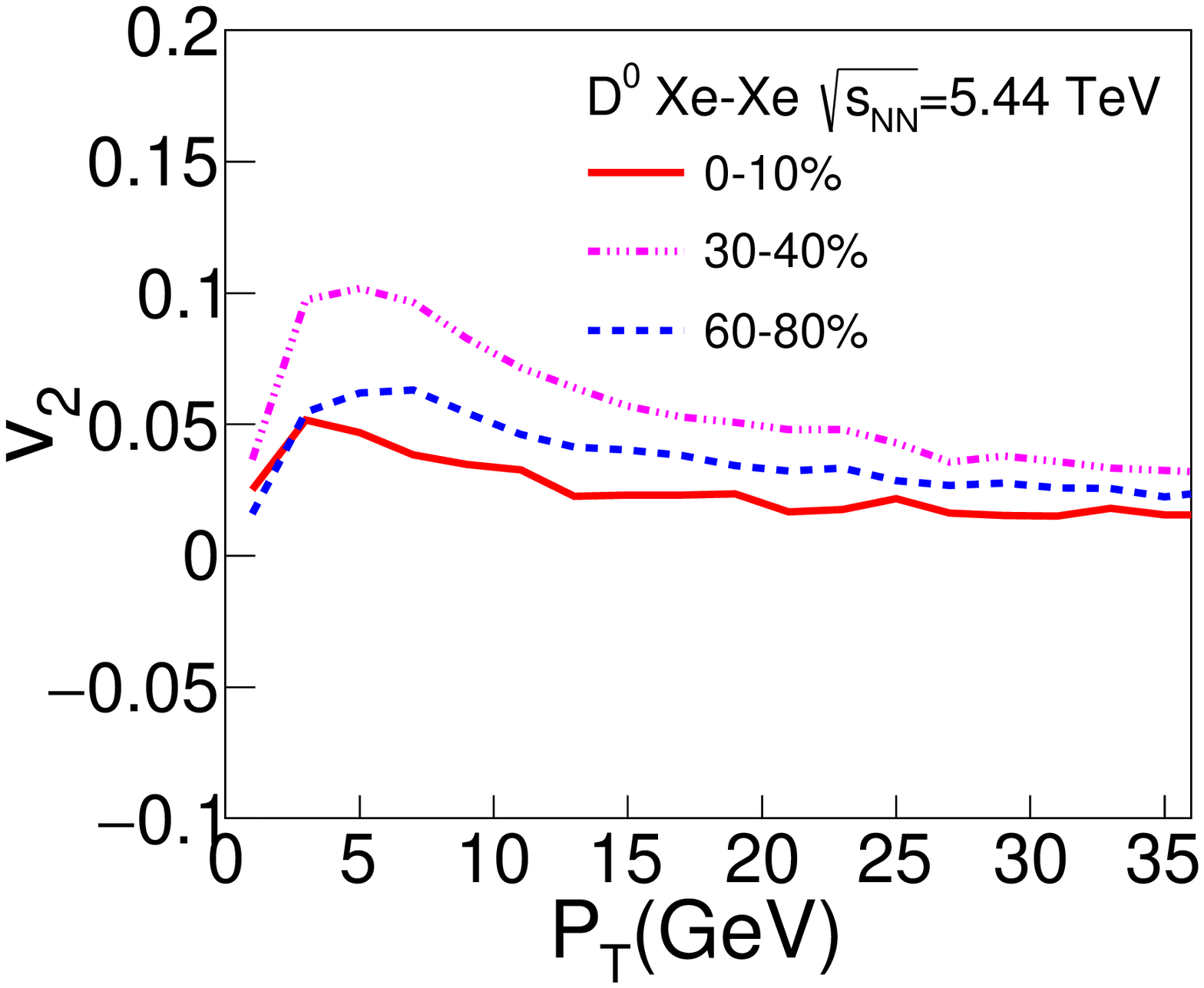}
    \includegraphics[clip=,width=0.25\textwidth]{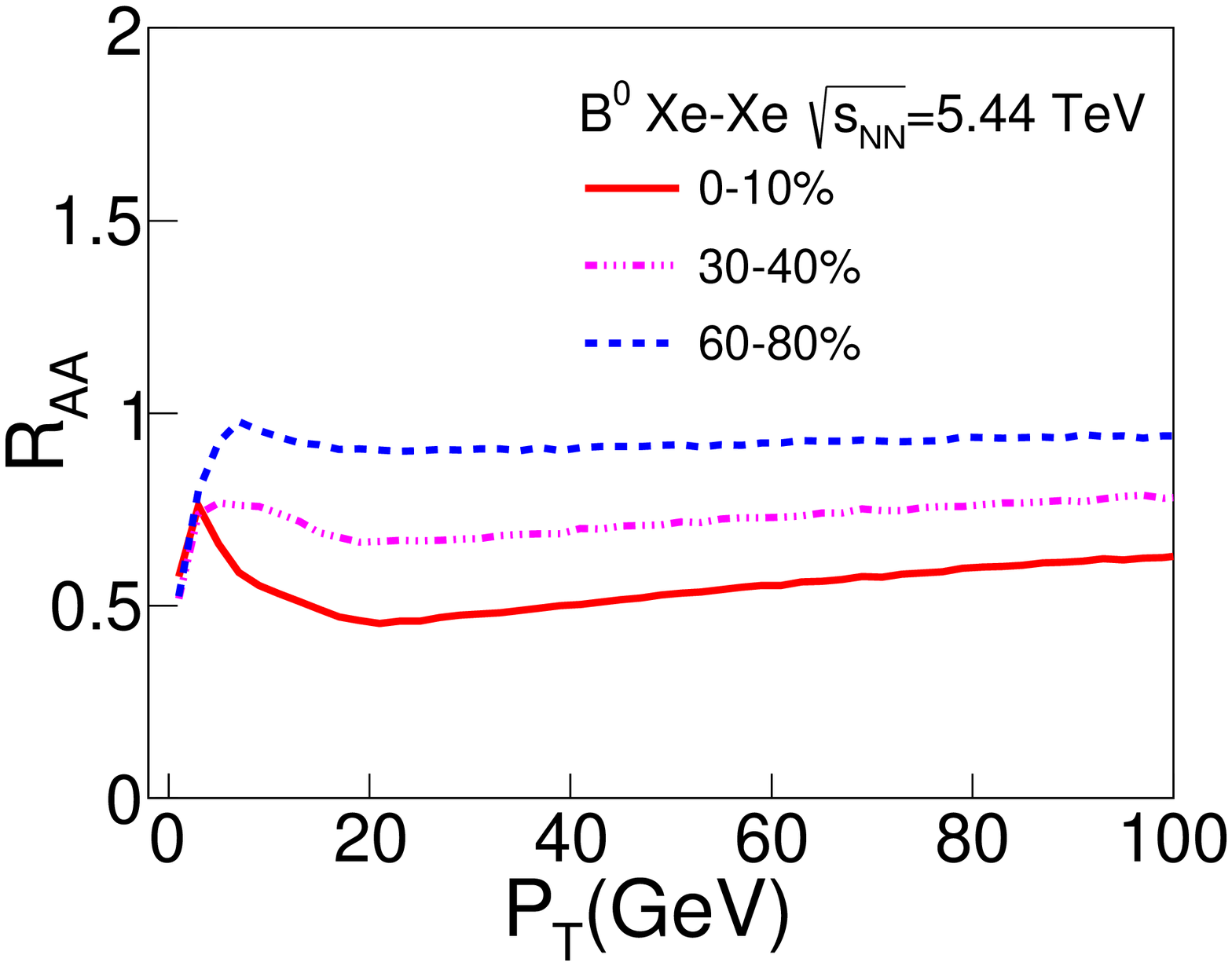}
    \hspace{-15pt}
    \includegraphics[clip=,width=0.25\textwidth]{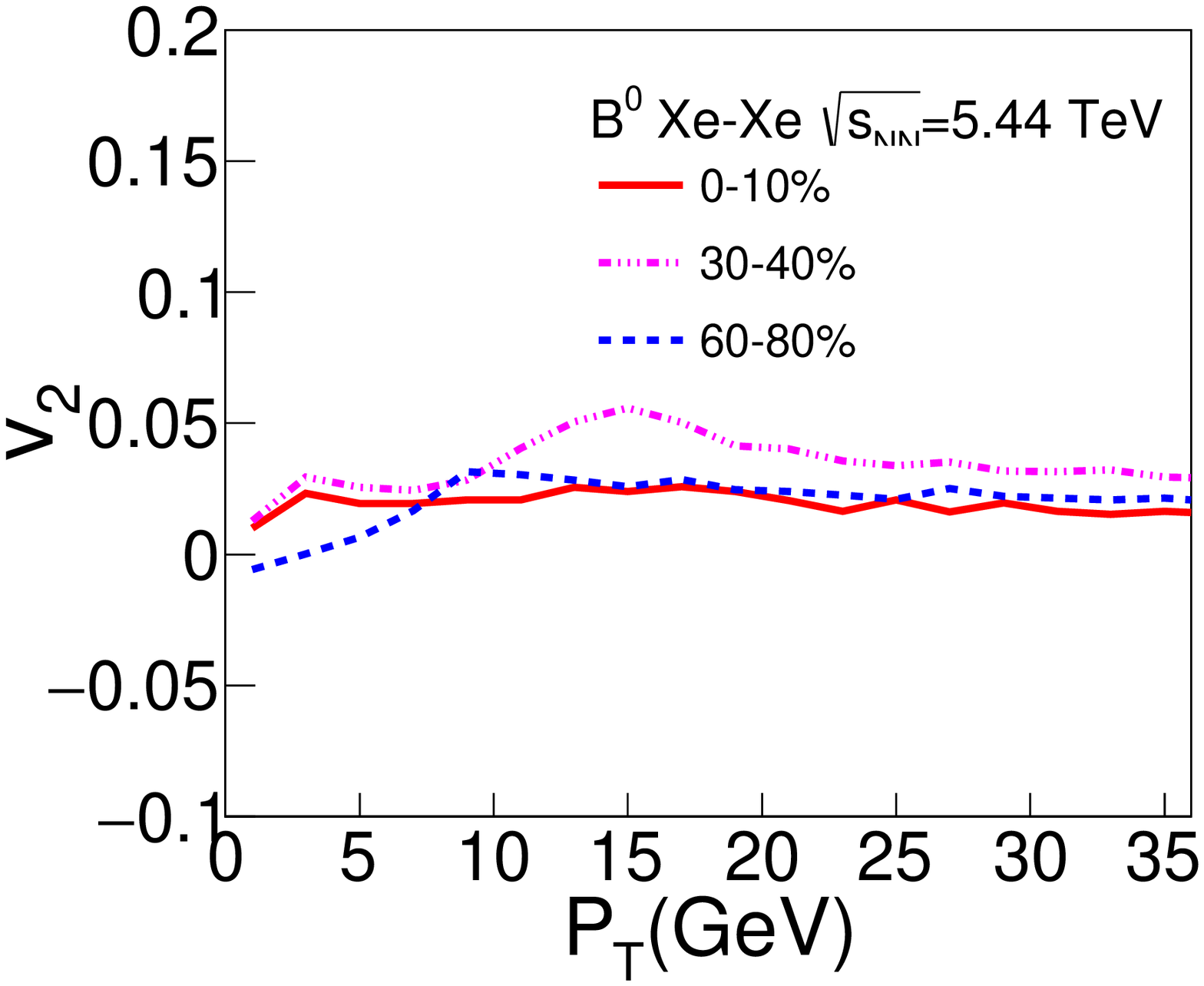}
    \caption{(Color online) Suppression (left) and elliptic flow coefficient (right) of $D$ mesons (upper) and $B$ mesons (lower) in different centrality classes of Xe-Xe collisions at $\sqrt{s_\mathrm{NN}}=5.44$~TeV.}
    \label{fig:raa-v2-pt-xexe}
\end{figure}

\begin{figure}[tbp]
    \centering
    \includegraphics[clip=,width=0.25\textwidth]{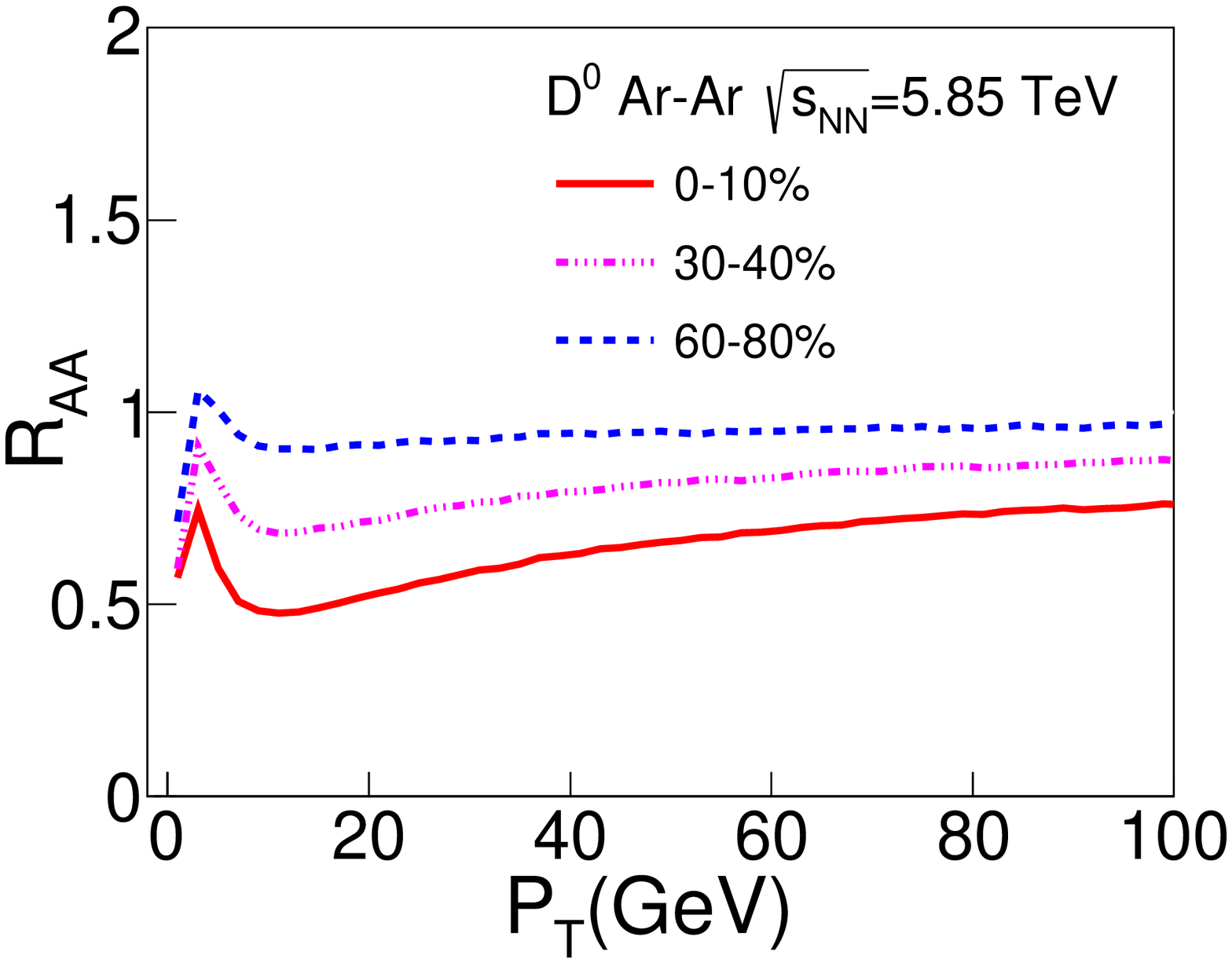}
    \hspace{-15pt}
    \includegraphics[clip=,width=0.25\textwidth]{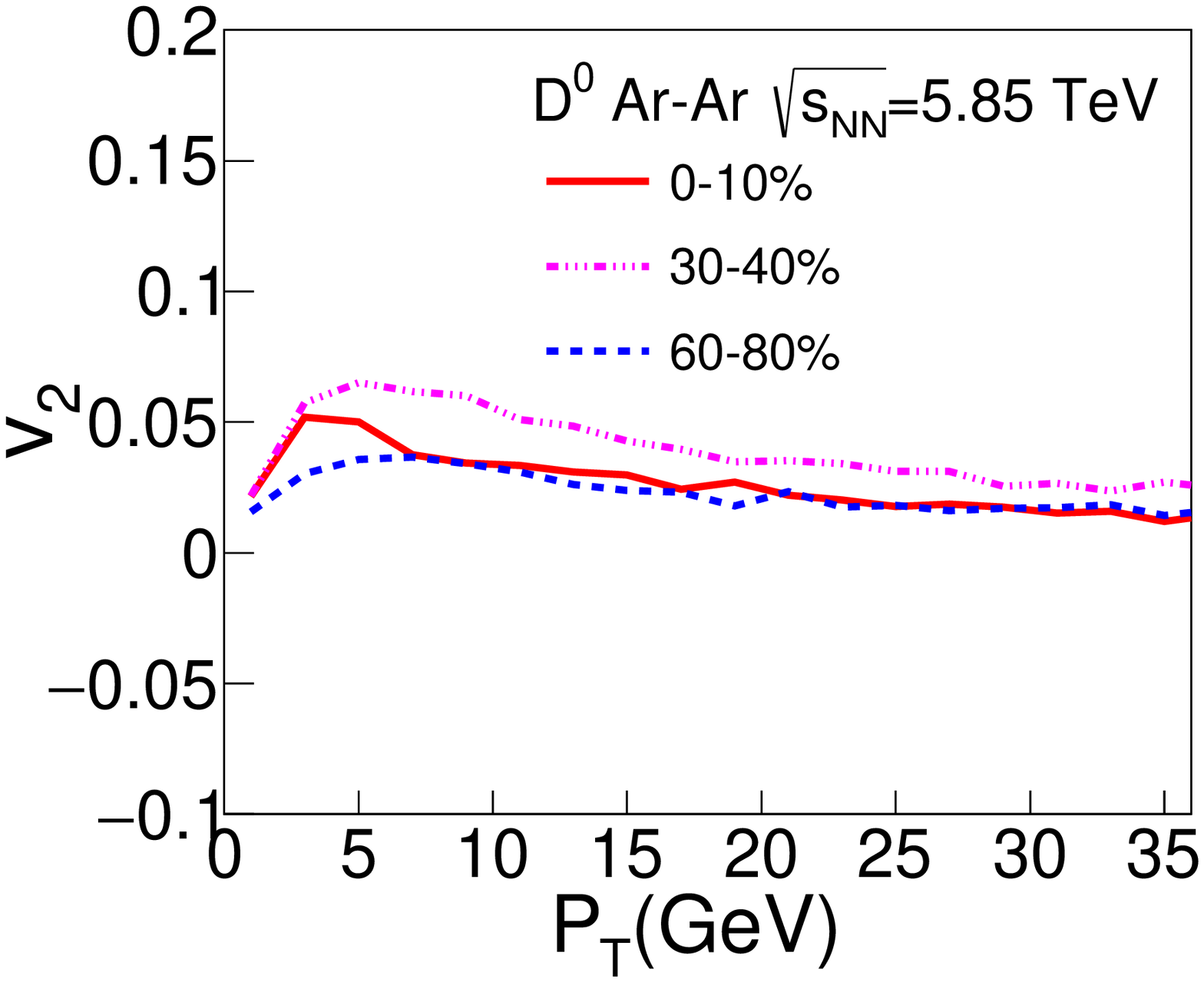}
    \includegraphics[clip=,width=0.25\textwidth]{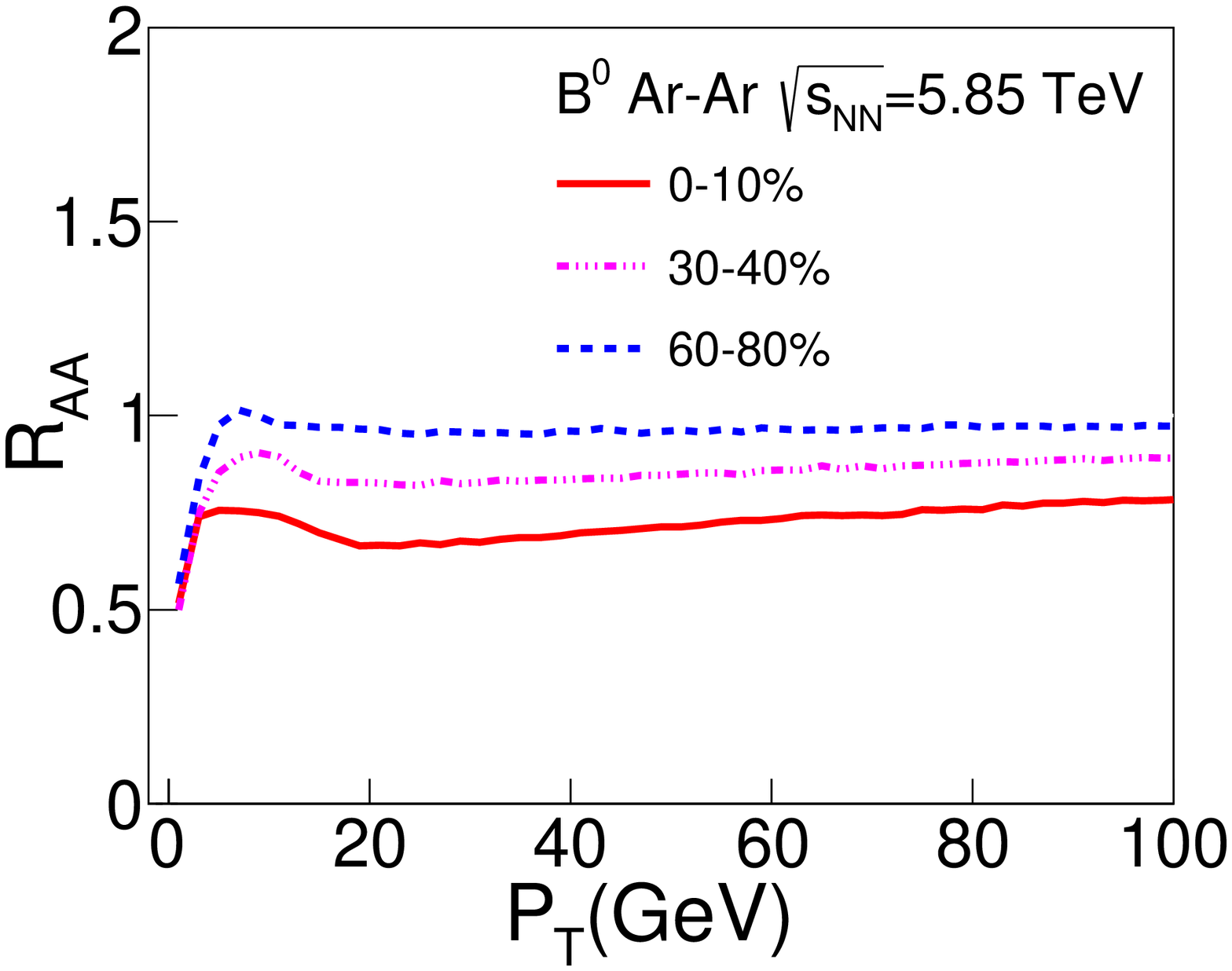}
    \hspace{-15pt}
    \includegraphics[clip=,width=0.25\textwidth]{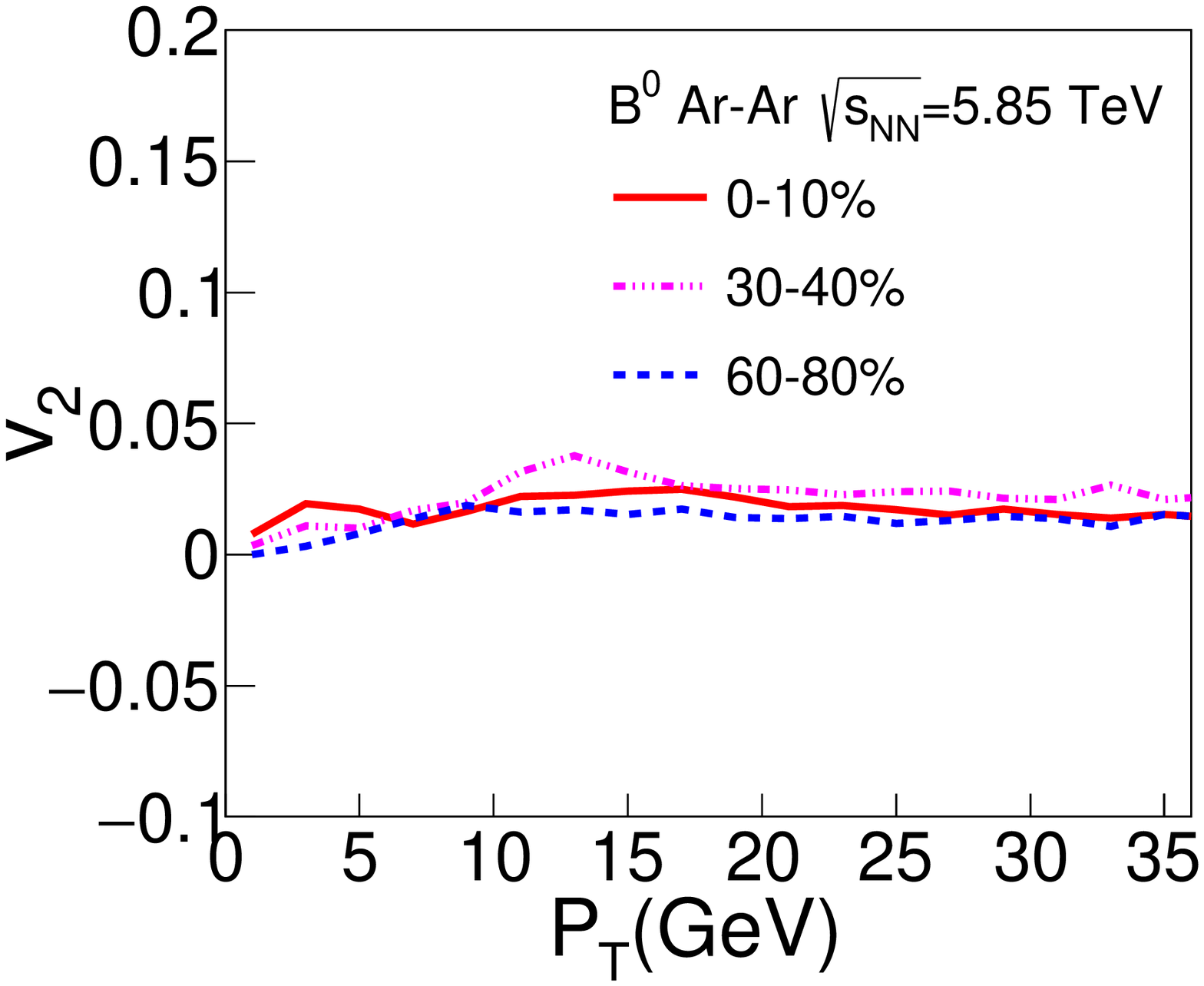}
    \caption{(Color online) Suppression (left) and elliptic flow coefficient (right) of $D$ mesons (upper) and $B$ mesons (lower) in different centrality classes of Ar-Ar collisions at $\sqrt{s_\mathrm{NN}}=5.85$~TeV.}
    \label{fig:raa-v2-pt-arar}
\end{figure}

\begin{figure}[tbp]
    \centering
    \includegraphics[clip=,width=0.25\textwidth]{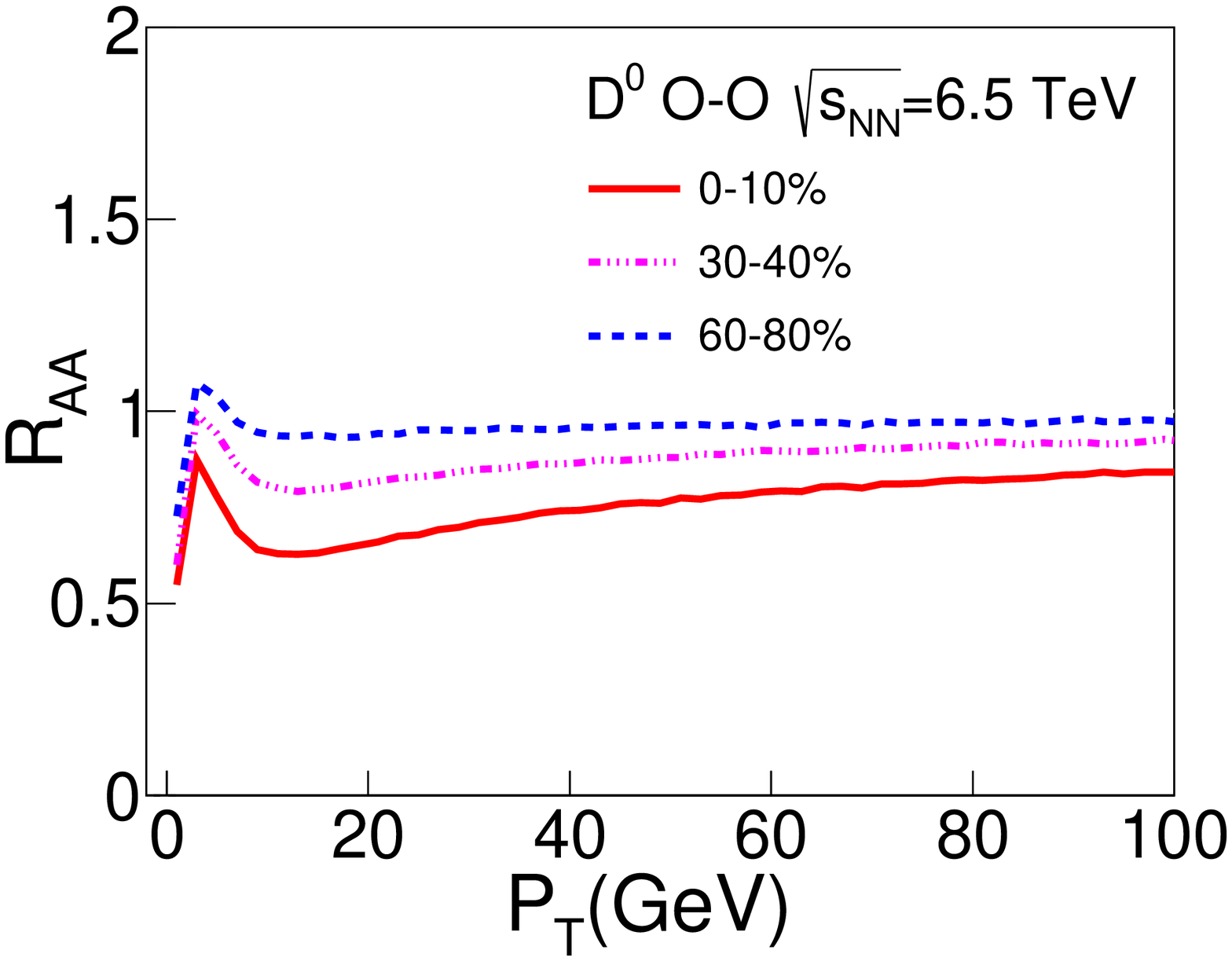}
    \hspace{-15pt}
    \includegraphics[clip=,width=0.25\textwidth]{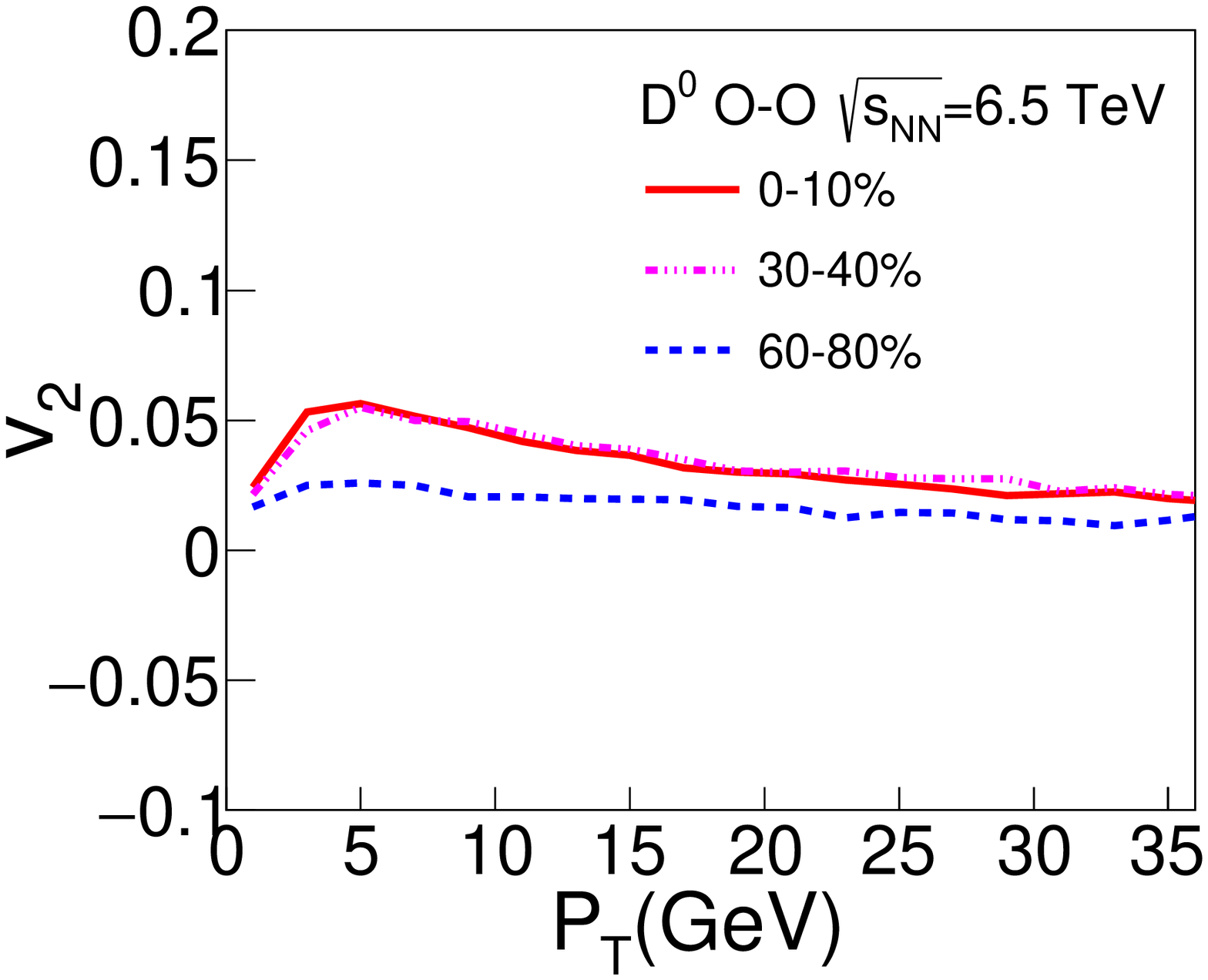}
    \includegraphics[clip=,width=0.25\textwidth]{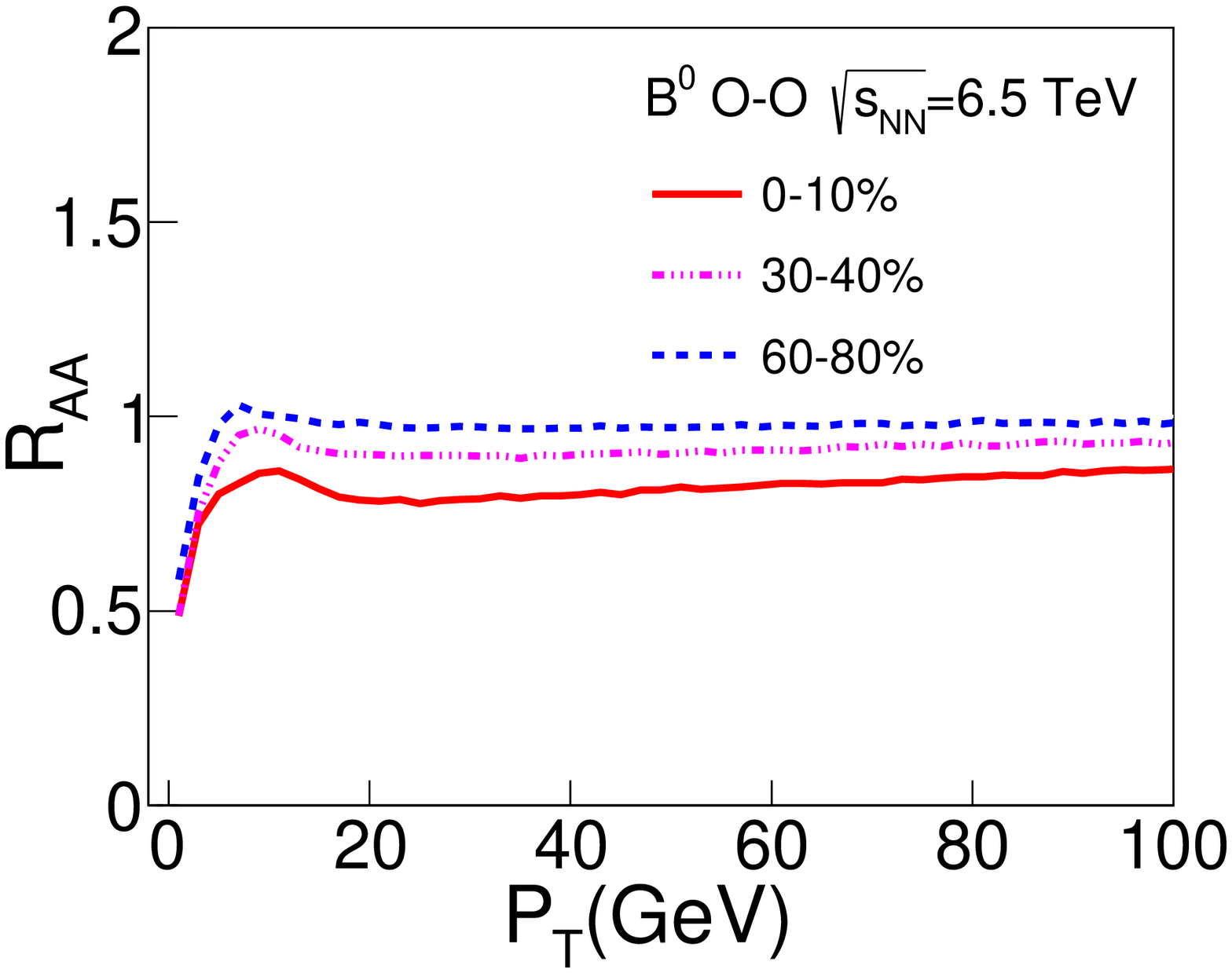}
    \hspace{-15pt}
    \includegraphics[clip=,width=0.25\textwidth]{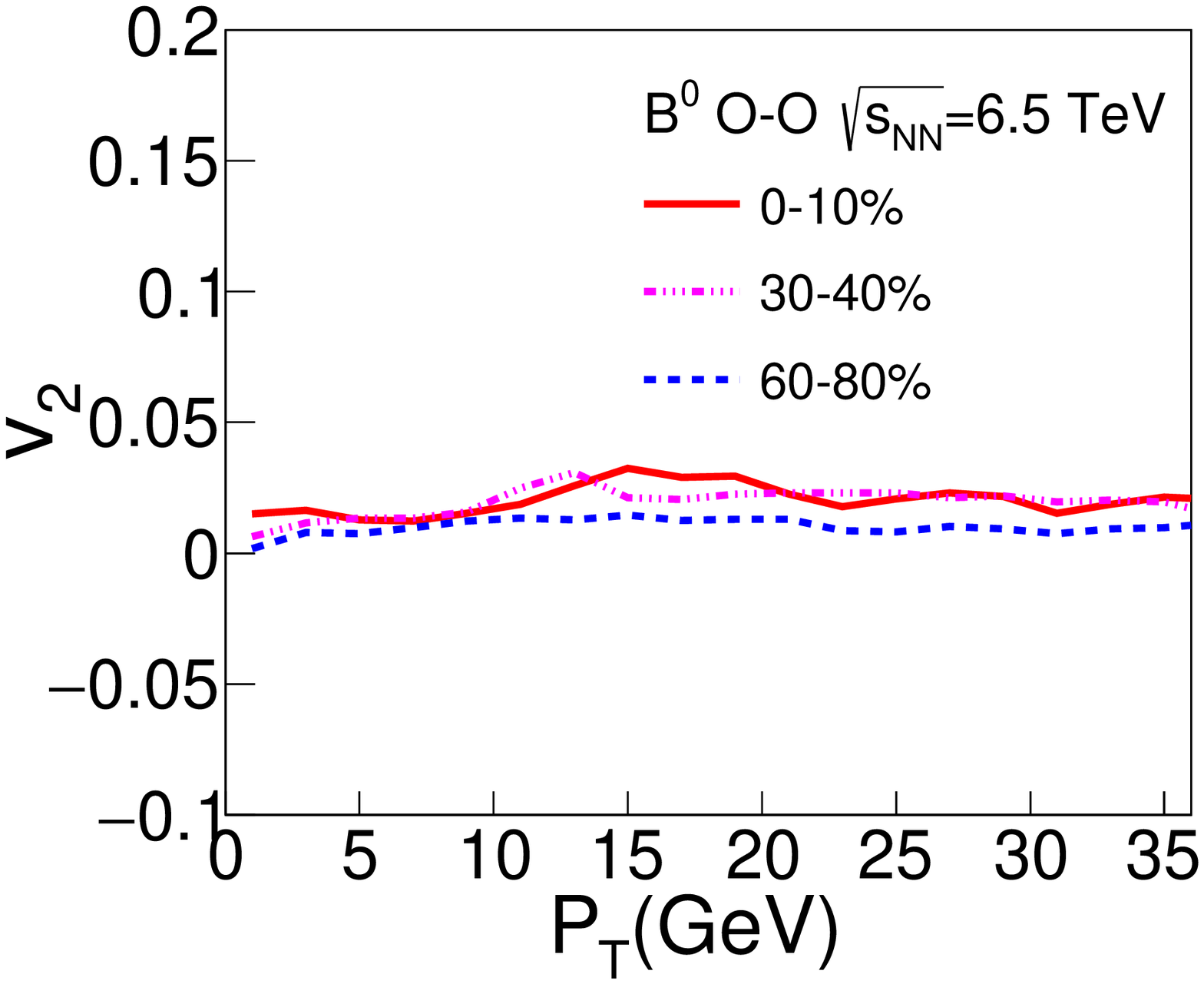}
    \caption{Suppression (left) and elliptic flow coefficient (right) of $D$ mesons (upper) and $B$ mesons (lower) in different centrality classes of O-O collisions at $\sqrt{s_\mathrm{NN}}=6.5$~TeV.}
    \label{fig:raa-v2-pt-oo}
\end{figure}

Within the same framework, we investigate the nuclear modification of heavy flavor mesons in smaller systems at the LHC. Results are presented in Fig.~\ref{fig:raa-v2-pt-xexe} for Xe-Xe collisions at $\sqrt{s_\mathrm{NN}}=5.44$~TeV, in Fig.~\ref{fig:raa-v2-pt-arar} for Ar-Ar collisions at $\sqrt{s_\mathrm{NN}}=5.85$~TeV, and in Fig.~\ref{fig:raa-v2-pt-oo} for O-O collisions at $\sqrt{s_\mathrm{NN}}=6.5$~TeV, in spite of the current absence of the corresponding experimental data. Similar to previous results for Pb-Pb collisions, in each figure, we present calculations for $D$ mesons in the upper panels and $B$ mesons in the lower panels, left for $R_\mathrm{AA}$ and right for $v_2$. In each panel, three centrality classes are compared, representing central (0-10\%), semi-central/peripheral (30-40\%) and peripheral (60-80\%) scenarios.

Comparing different collision systems (from Fig.~\ref{fig:raa-v2-pt-pbpb} to Fig.~\ref{fig:raa-v2-pt-oo}), a general conclusion can be drawn, i.e., parton energy loss becomes weaker inside a smaller collision system, as suggested by the gradually larger heavy flavor meson $R_\mathrm{AA}$ and smaller $v_2$ within the same centrality class as we move from Pb-Pb, Xe-Xe, Ar-Ar to O-O collisions. Such system size dependence of jet quenching effects provides a crucial bridge of jet-medium interaction between large and small collision systems.

It is interesting to note that even in the relatively small system produced by O-O collisions, considerable amount of energy loss effects are found for both charm and bottom quarks in the most central collisions -- the corresponding $D$ and $B$ meson $R_\mathrm{AA}$'s are significantly smaller than unity while their $v_2$'s have sizable values. As one moves from central to peripheral collisions, heavy flavor meson $R_\mathrm{AA}$ increases and approaches unity at high $p_\mathrm{T}$ in peripheral collisions. The rise-and-fall structure of $R_{\rm AA}$ at low $p_\mathrm{T}$ region is due to the coalescence mechanism in heavy hadron formation in the presence of QGP medium, which converts low $p_\mathrm{T}$ heavy quarks into intermediate $p_\mathrm{T}$ heavy flavor mesons. For the heavy flavor meson $v_2$, it first increases and then decreases as a function of centrality class due to the competing effects between parton energy loss and geometric anisotropy of the collision zone.

\begin{figure}[tbp]
    \centering
    \includegraphics[clip=,width=0.25\textwidth]{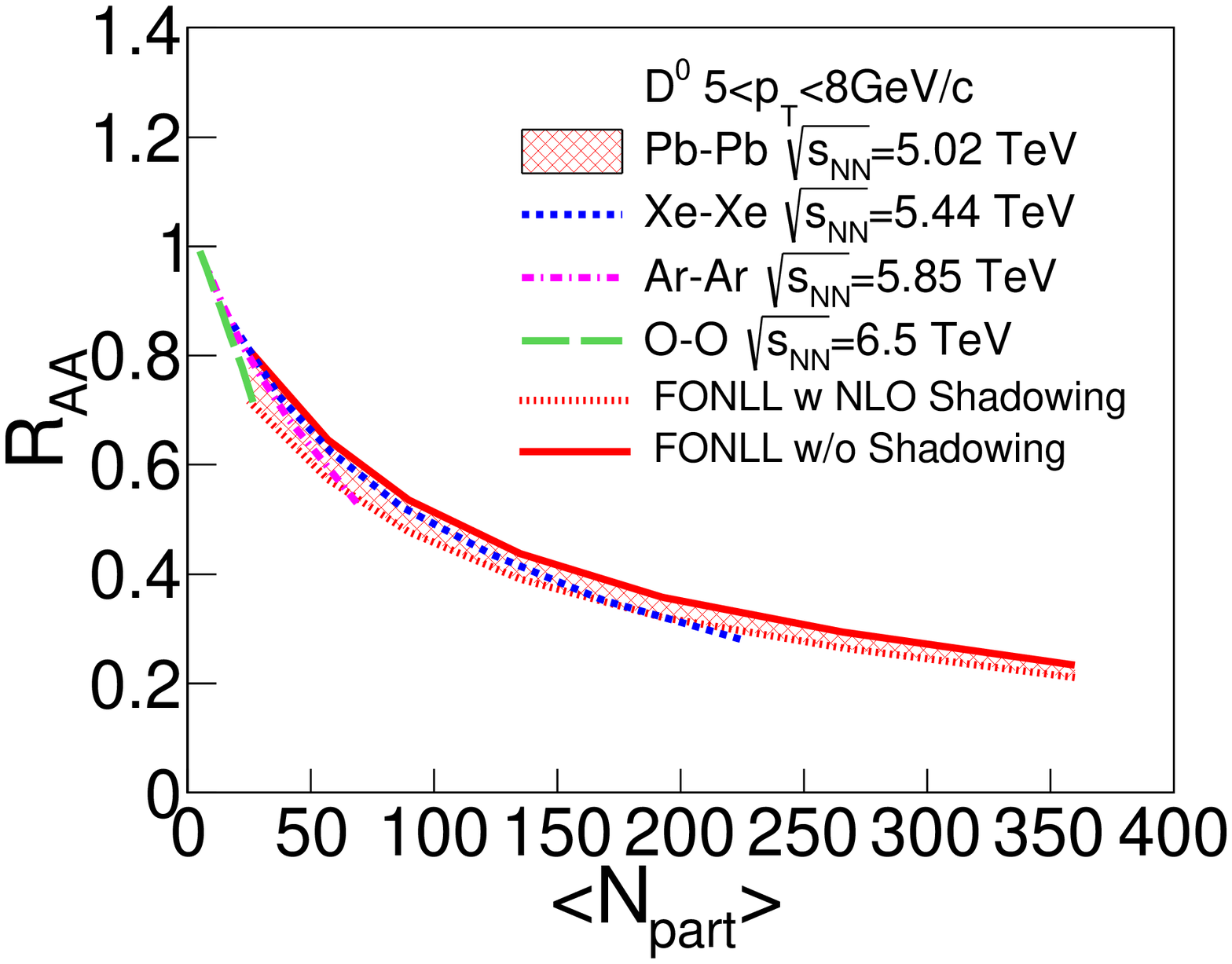}
    \hspace{-15pt}
    \includegraphics[clip=,width=0.25\textwidth]{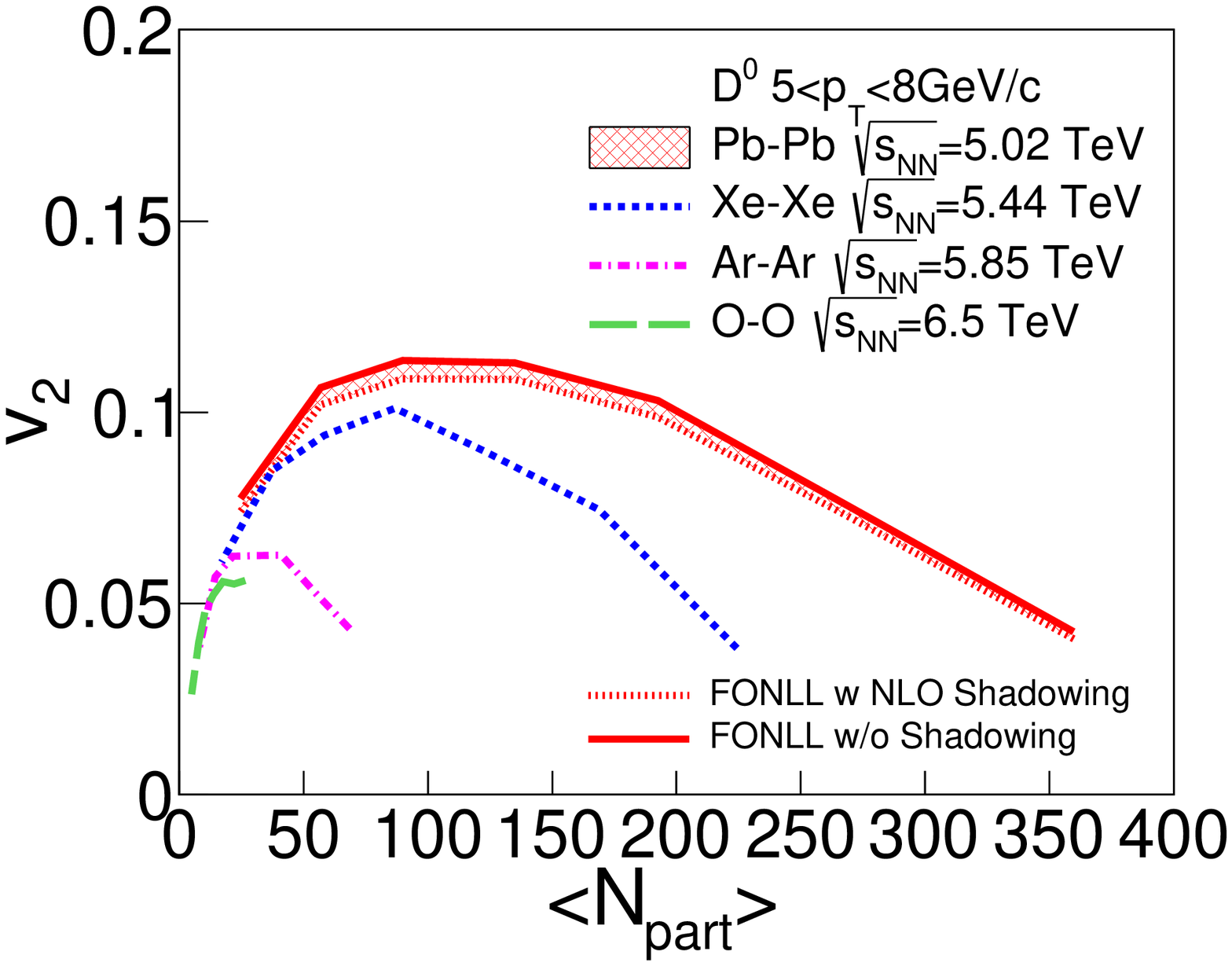}
    \includegraphics[clip=,width=0.25\textwidth]{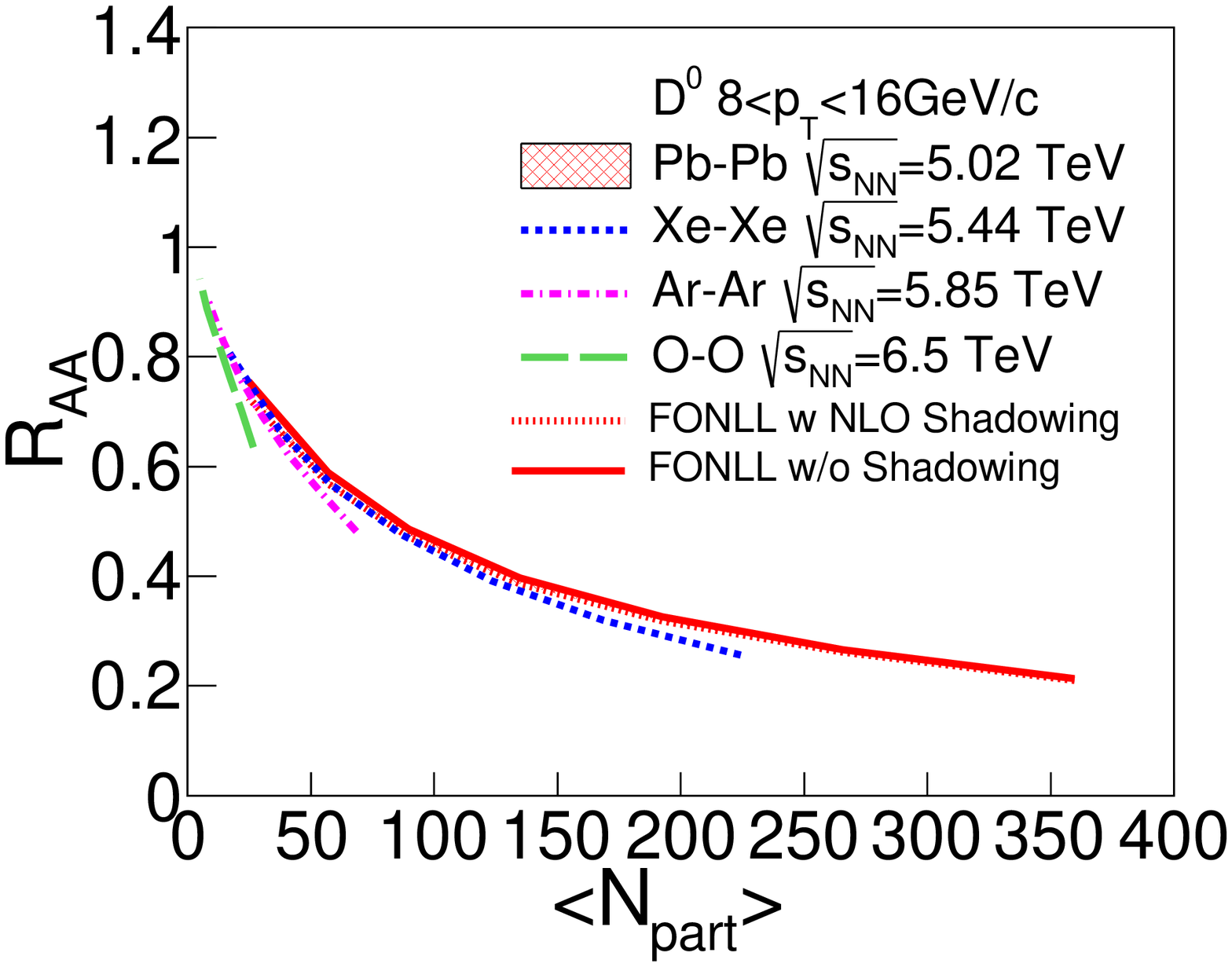}
    \hspace{-15pt}
    \includegraphics[clip=,width=0.25\textwidth]{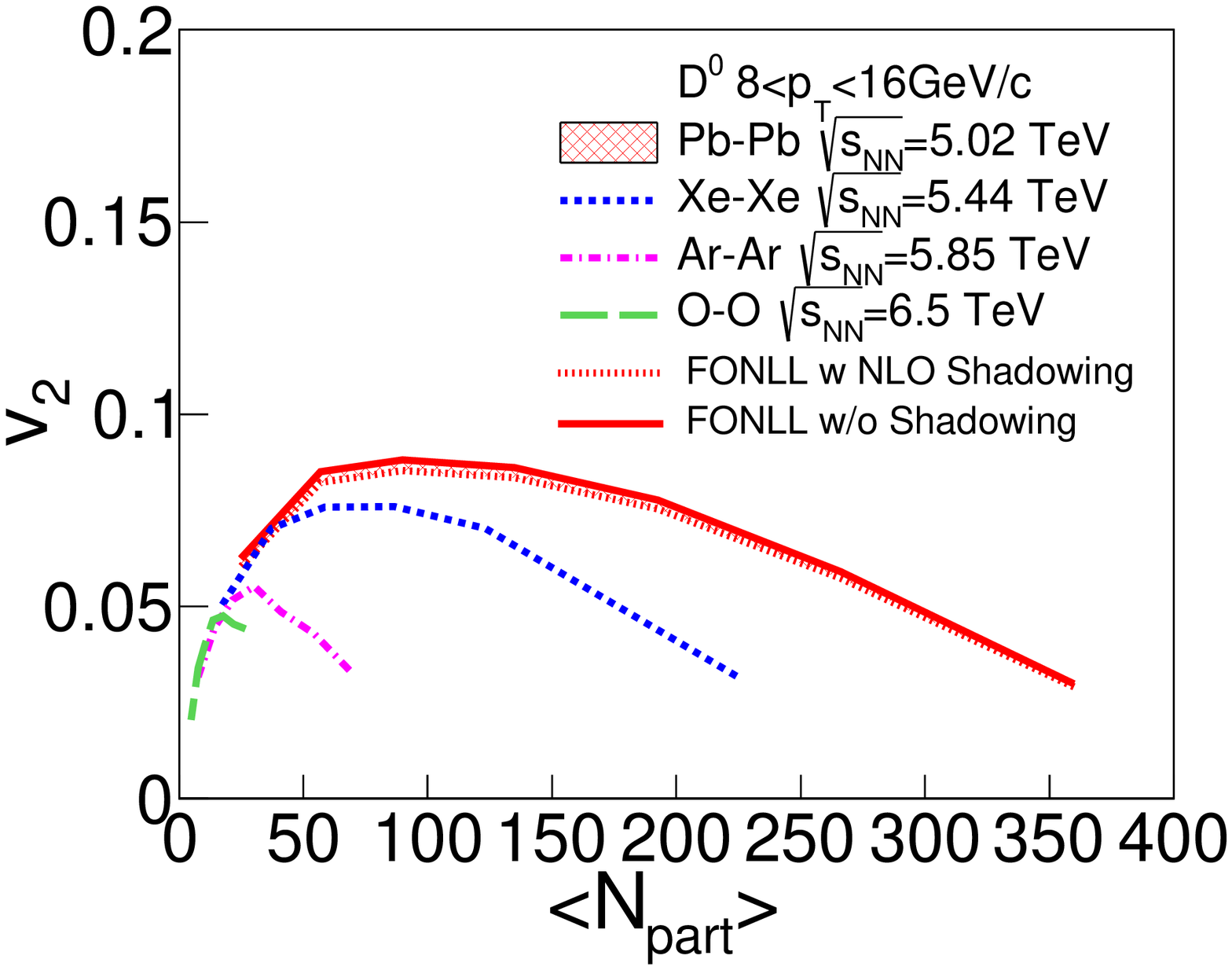}
    \caption{(Color online) Participant number dependence of $R_\mathrm{AA}$ (left) and $v_2$ (right) of $D$ mesons in different collision systems, upper for $5<p_\mathrm{T}<8$~GeV and lower for $8<p_\mathrm{T}<16$~GeV.}
    \label{fig:raa-v2-npart-D0}
\end{figure}

\begin{figure}[tbp]
    \centering
    \includegraphics[clip=,width=0.25\textwidth]{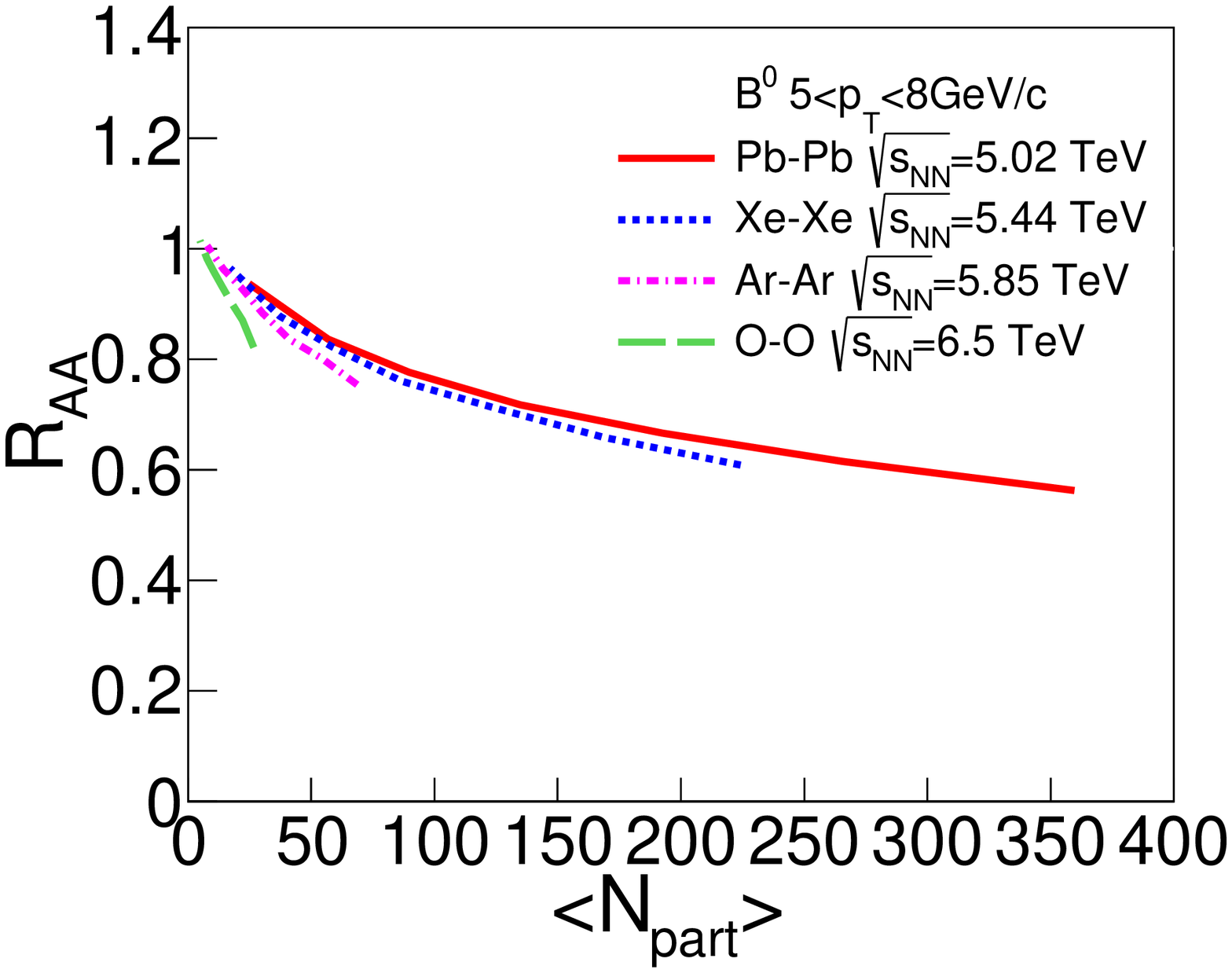}
    \hspace{-15pt}
    \includegraphics[clip=,width=0.25\textwidth]{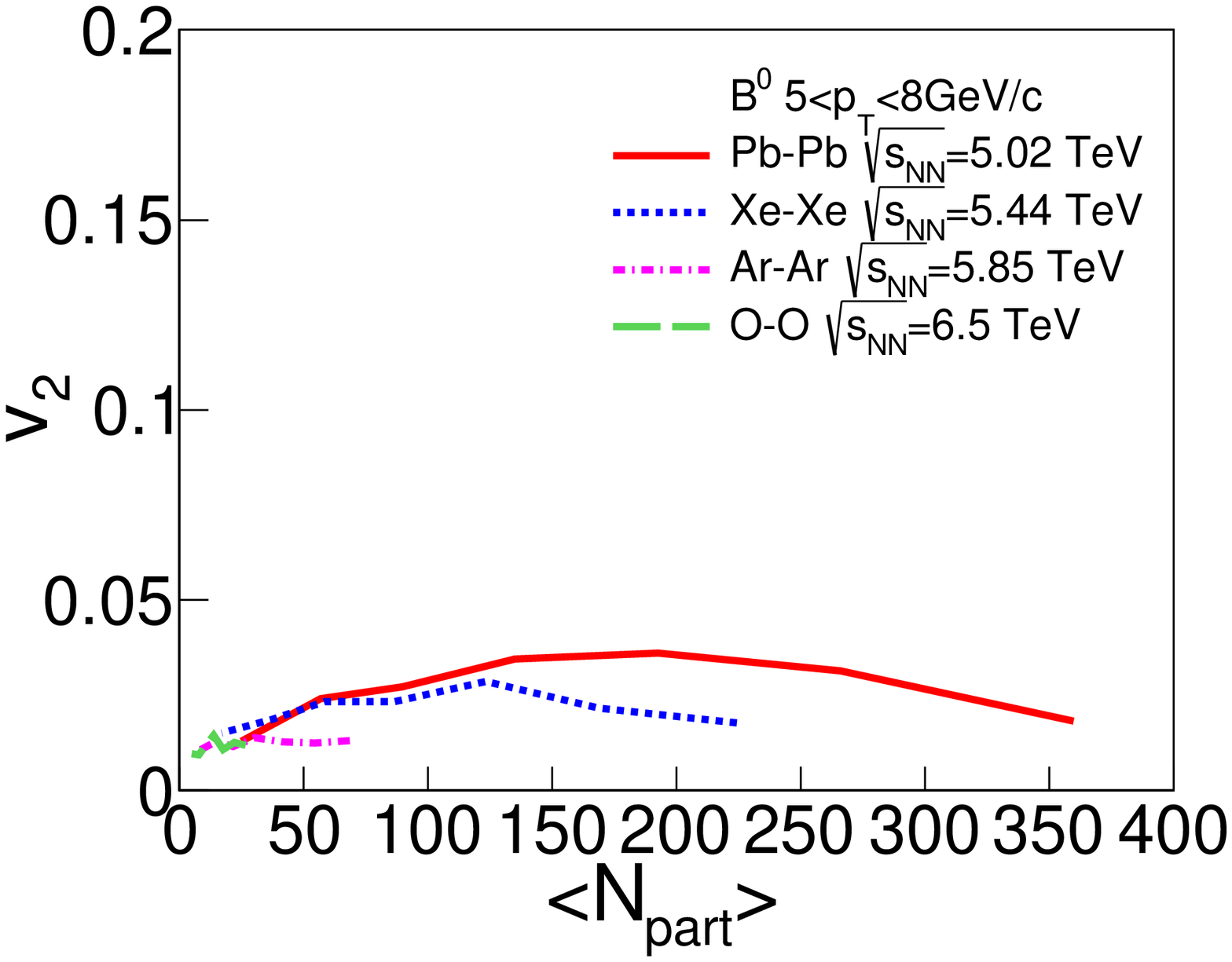}
    \includegraphics[clip=,width=0.25\textwidth]{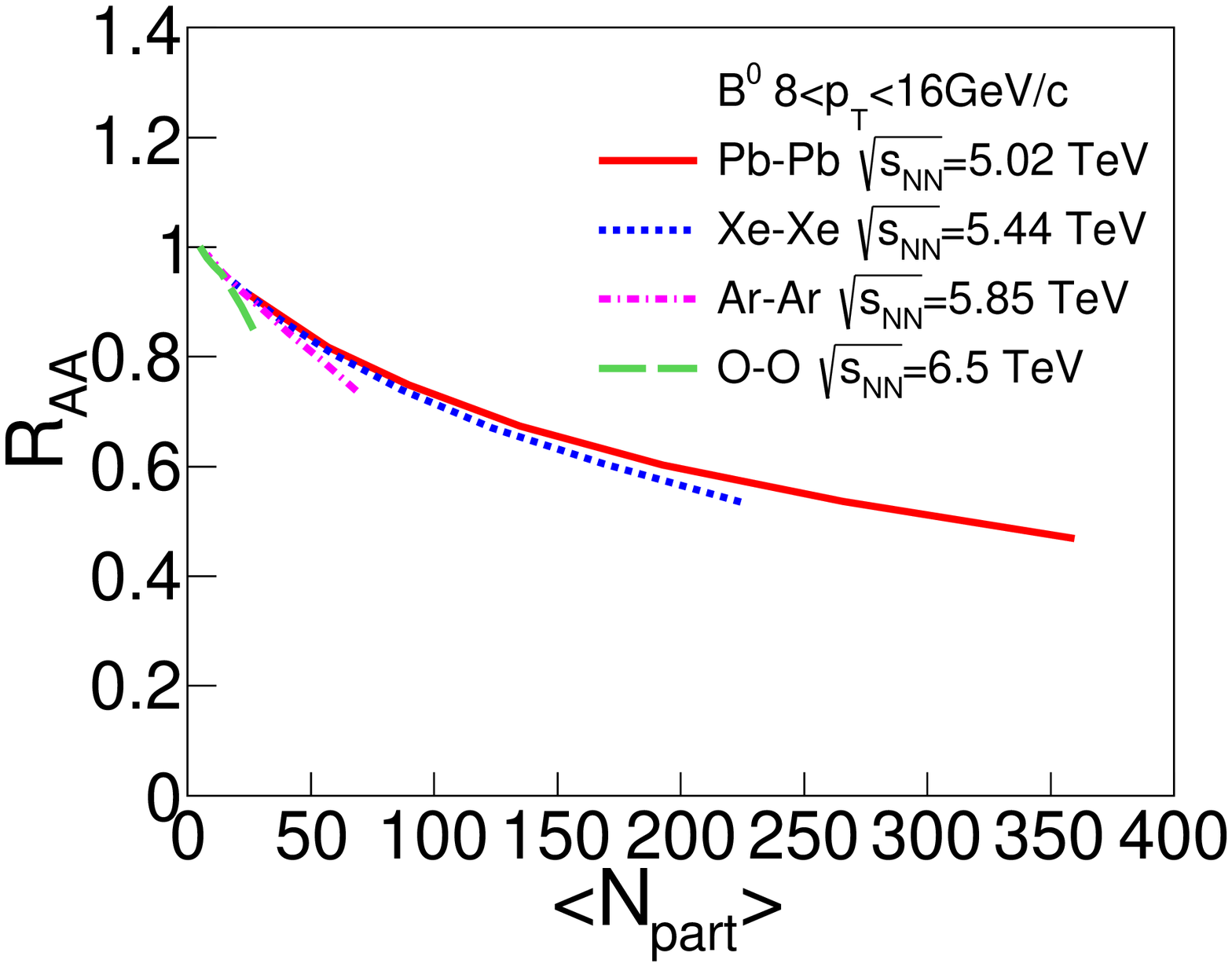}
    \hspace{-15pt}
    \includegraphics[clip=,width=0.25\textwidth]{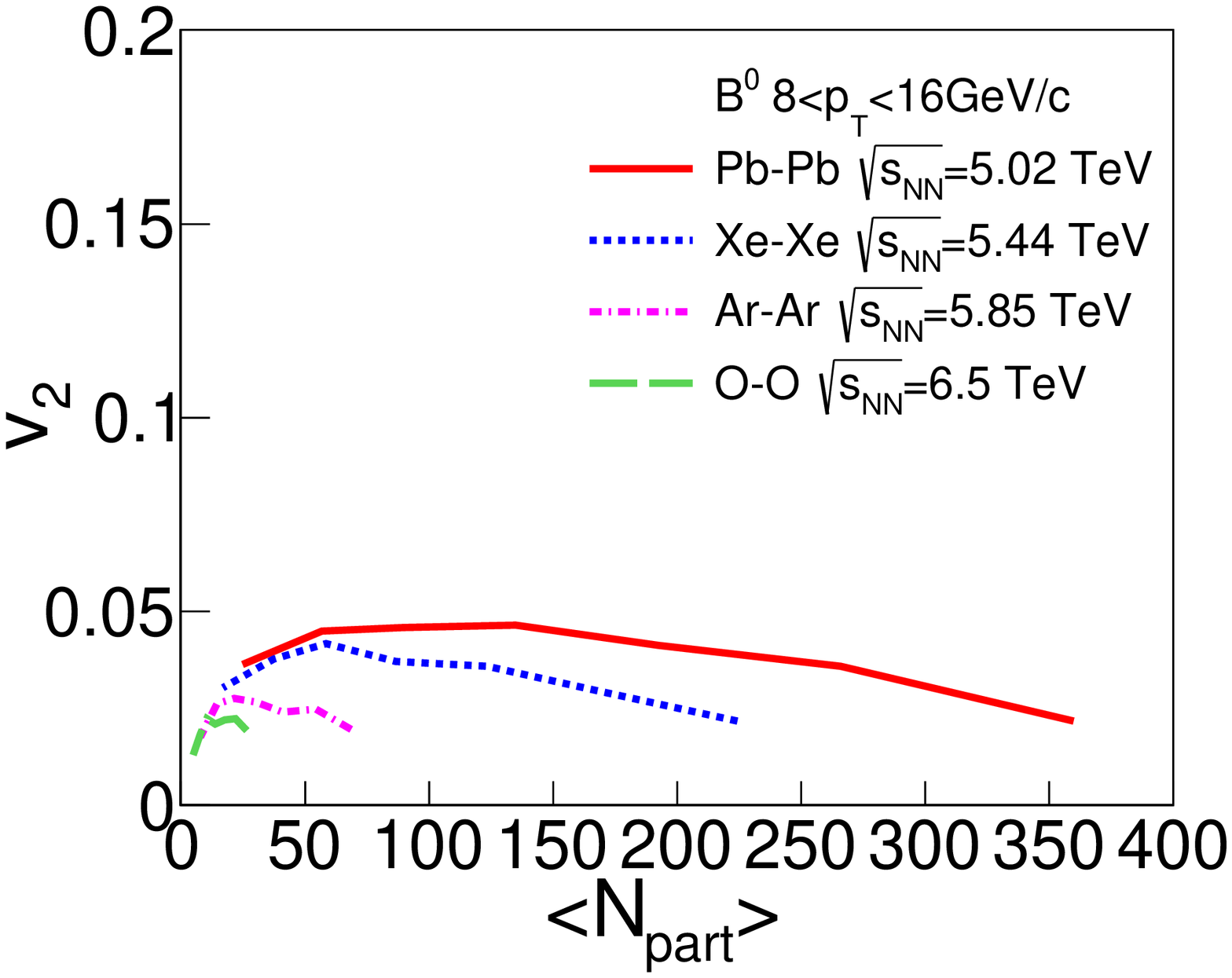}
    \caption{(Color online) Participant number dependence of $R_\mathrm{AA}$ (left) and $v_2$ (right) of $B$ mesons in different collision systems, upper for $5<p_\mathrm{T}<8$~GeV and lower for $8<p_\mathrm{T}<16$~GeV.}
    \label{fig:raa-v2-npart-B0}
\end{figure}

To have a more quantitative understanding of how heavy quark energy loss depends on the system size of QGP, we present the participant number ($N_\mathrm{part}$) dependence of the $R_\mathrm{AA}$ and $v_2$ of $D$ mesons in Fig.~\ref{fig:raa-v2-npart-D0} and $B$ mesons in Fig.~\ref{fig:raa-v2-npart-B0}. In each panel of these two figures, we present the $p_\mathrm{T}$-integrated observable as a function of $N_\mathrm{part}$ for different collision systems.
The upper panels are for $5<p_\mathrm{T}<8$~GeV and the lower for $8<p_\mathrm{T}<16$~GeV. In the left panels, we observe a stronger nuclear modification of heavy mesons with a larger $N_\mathrm{part}$. As previously discussed, one can find clear nuclear modification of both $D$ and $B$ mesons even in the small-size O-O collisions as long as $N_\mathrm{part}$ is not small. In addition, a scaling behavior of the nuclear modification factor with respect to $N_\mathrm{part}$ can be seen in the left panels: the heavy flavor meson $R_\mathrm{AA}$ in different collision systems follow the similar $N_\mathrm{part}$ dependence. In other words, heavy flavor mesons produced from different collision systems share a similar $R_\mathrm{AA}$ as long as $N_\mathrm{part}$ is fixed. The slight breaking of this $N_\mathrm{part}$ scaling behavior shown in the figures could be due to different initial heavy quark spectra produced at different $\sqrt{s_\mathrm{NN}}$ for different collision systems.

Unlike $R_\mathrm{AA}$, the $N_\mathrm{part}$ scaling behavior does not exist for $v_2$, as shown in the right panels of Figs.~\ref{fig:raa-v2-npart-D0} and~\ref{fig:raa-v2-npart-B0}. This is because $v_2$ is driven not only by the overall energy loss of heavy quarks that is determined by $N_\mathrm{part}$, but also by the geometric anisotropy of the medium. For the same centrality class, larger collision system (e.g. Pb-Pb) has higher $N_\mathrm{part}$ than smaller system (e.g. O-O). In other words, for similar $N_\mathrm{part}$, larger system has stronger anisotropy. Therefore, one observes the hierarchy of Pb-Pb $>$ Xe-Xe $>$ Ar-Ar $>$ O-O for the heavy meson $v_2$ in the right panels of these two figures.

\begin{figure}[tbp]
    \centering
    \includegraphics[clip=,width=0.25\textwidth]{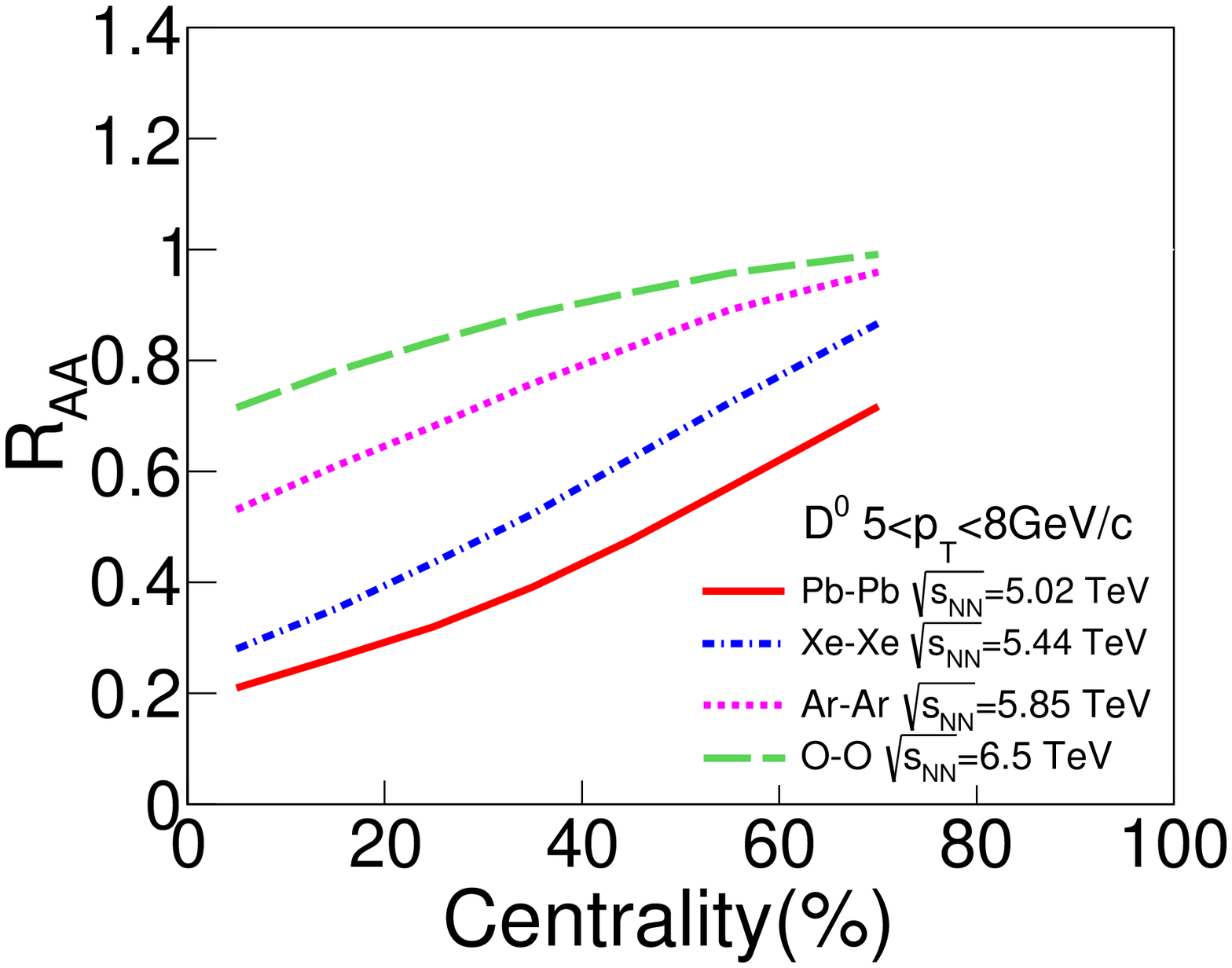}
    \hspace{-15pt}
    \includegraphics[clip=,width=0.25\textwidth]{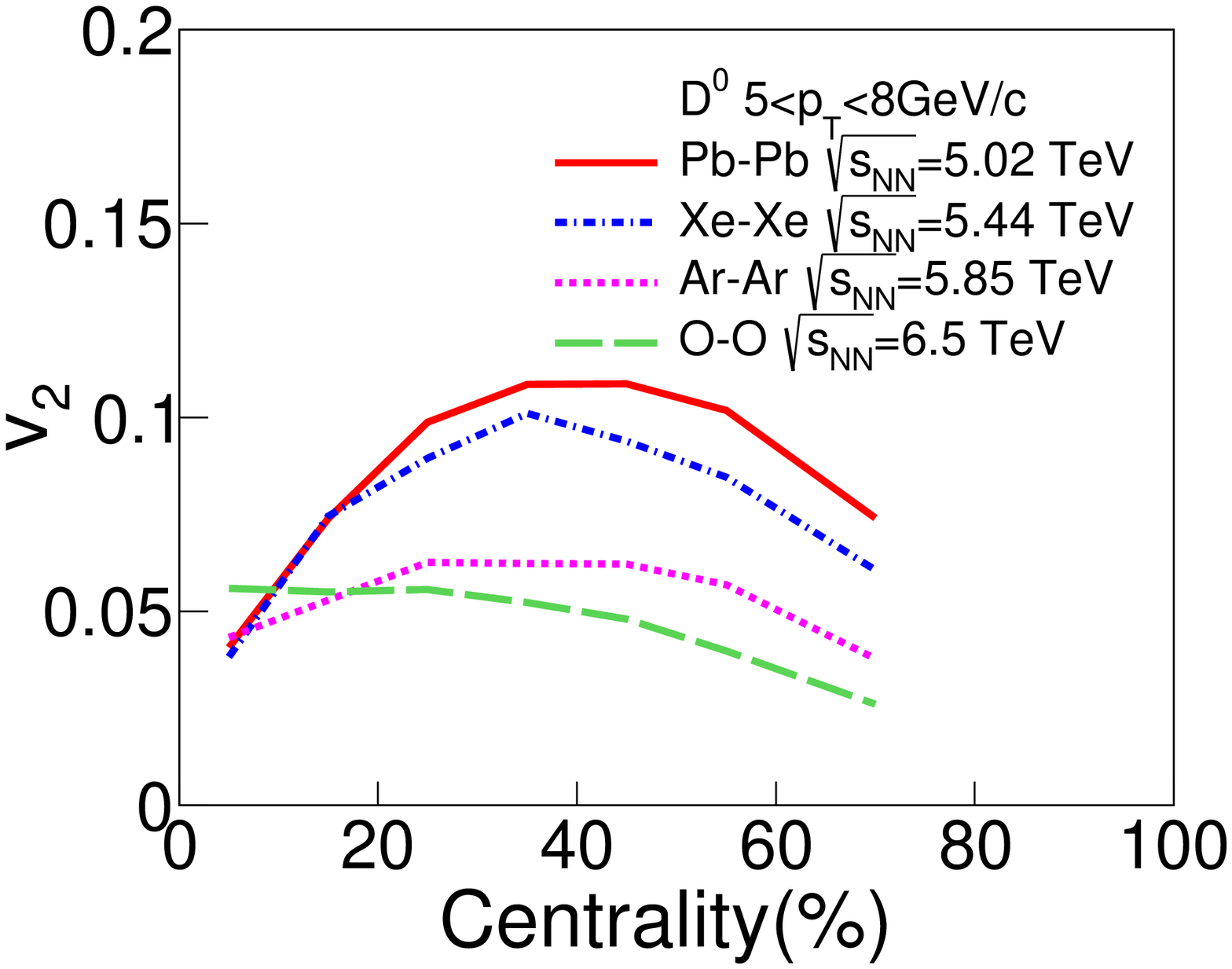}
    \includegraphics[clip=,width=0.25\textwidth]{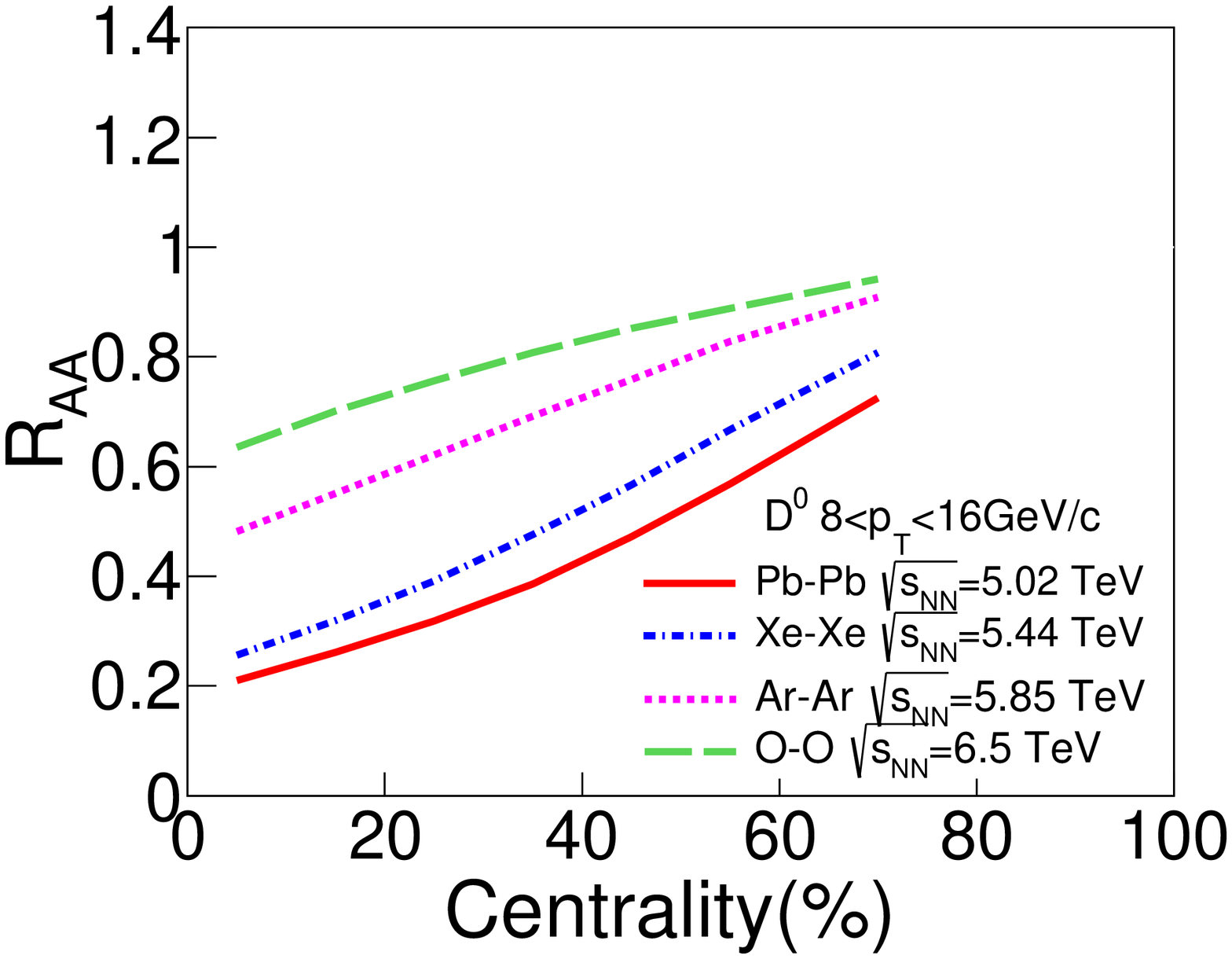}
    \hspace{-15pt}
    \includegraphics[clip=,width=0.25\textwidth]{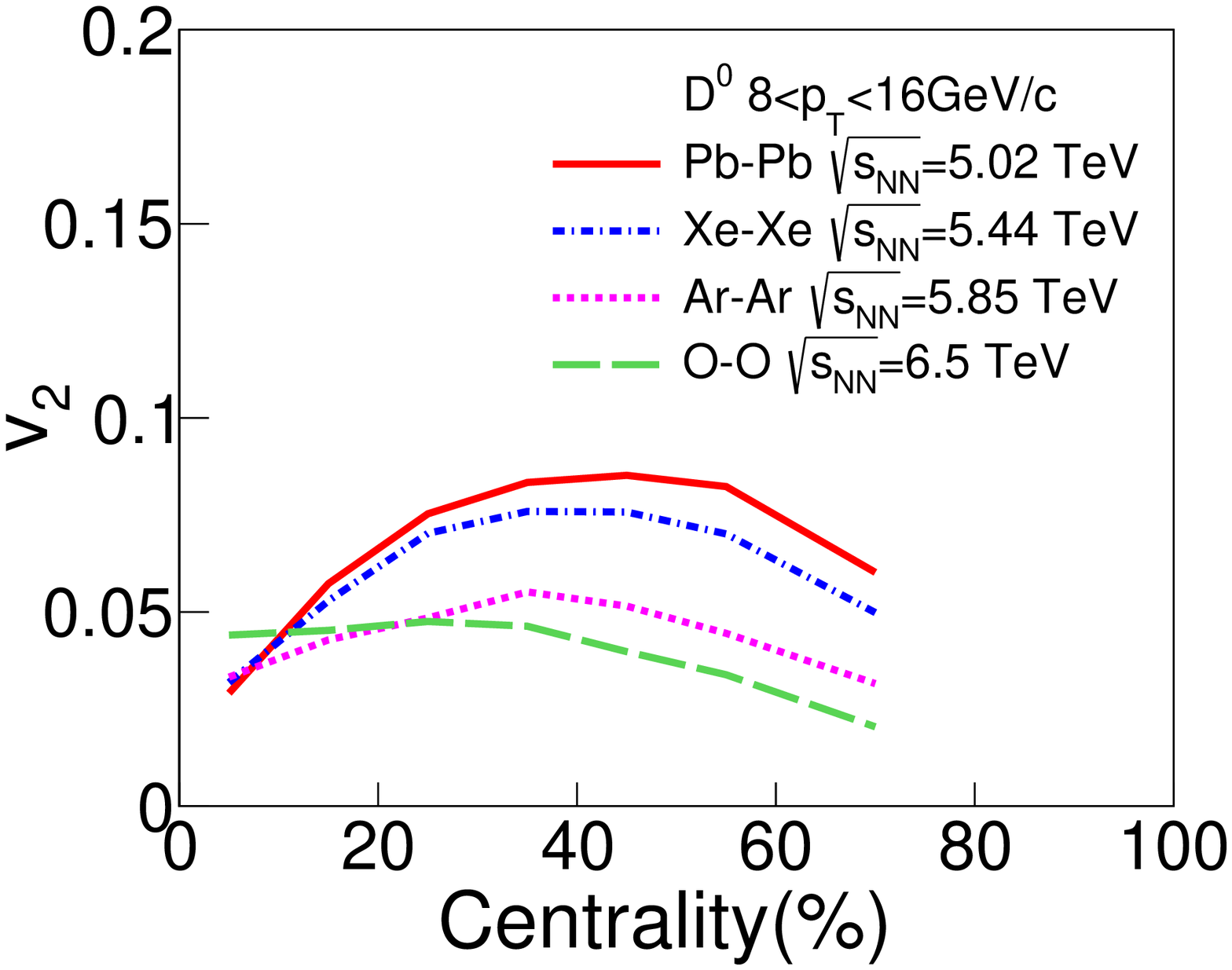}
    \caption{(Color online) Centrality dependence of $R_\mathrm{AA}$ (left) and $v_2$ (right) of $D$ mesons in different collision systems, upper for $5<p_\mathrm{T}<8$~GeV and lower for $8<p_\mathrm{T}<16$~GeV.}
    \label{fig:raa-v2-centrality-D0}
\end{figure}

\begin{figure}[tbp]
    \centering
    \includegraphics[clip=,width=0.25\textwidth]{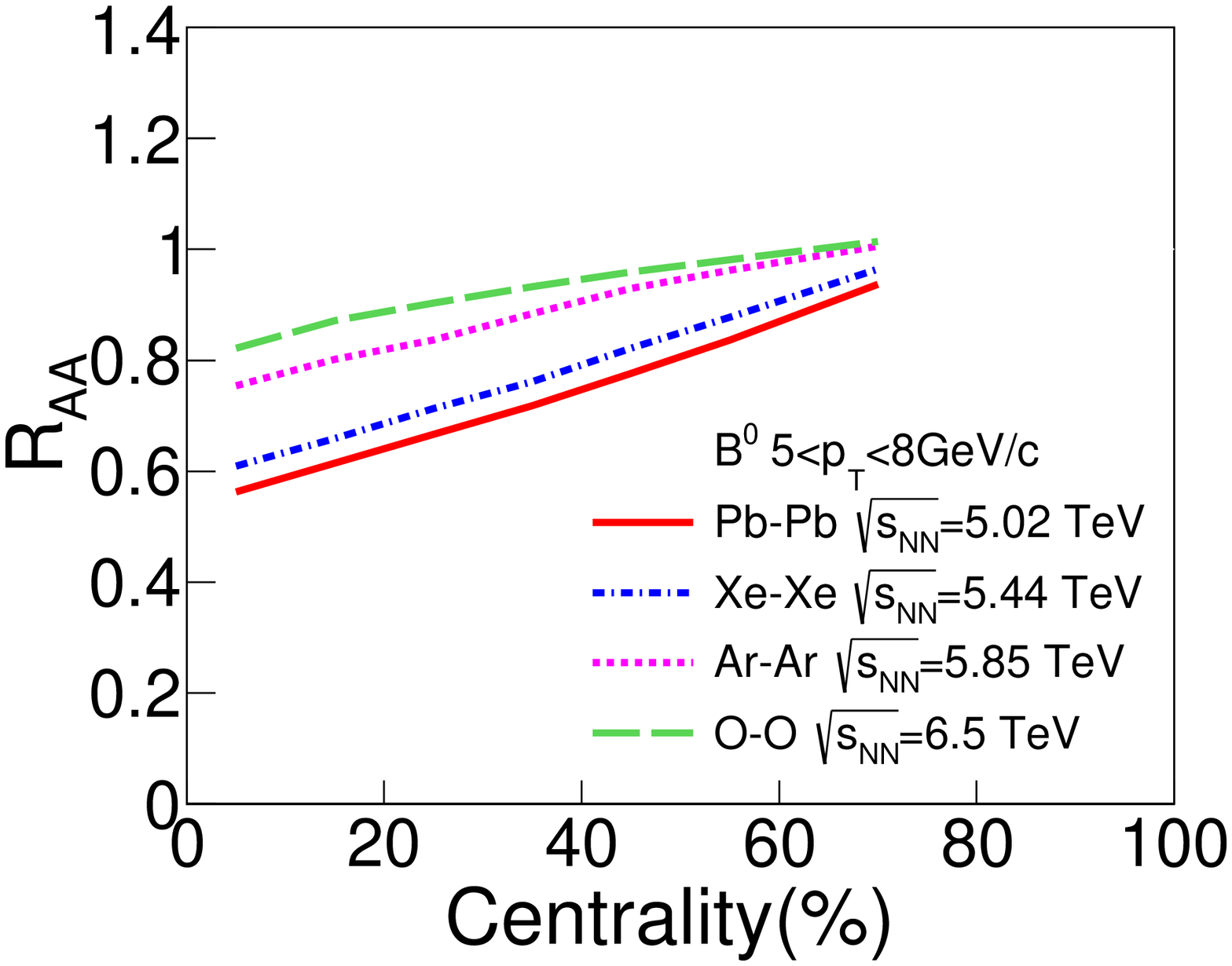}
    \hspace{-15pt}
    \includegraphics[clip=,width=0.25\textwidth]{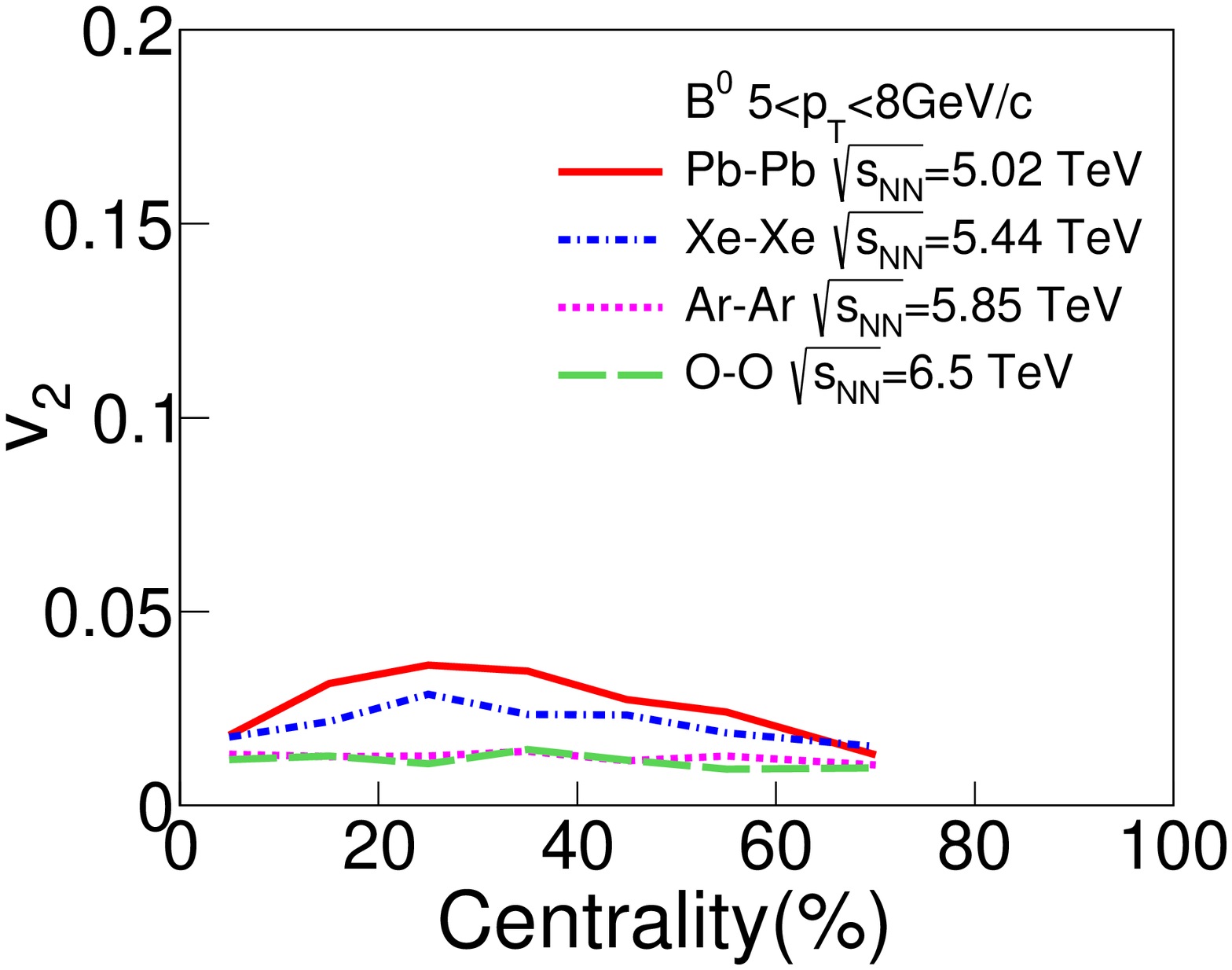}
    \includegraphics[clip=,width=0.25\textwidth]{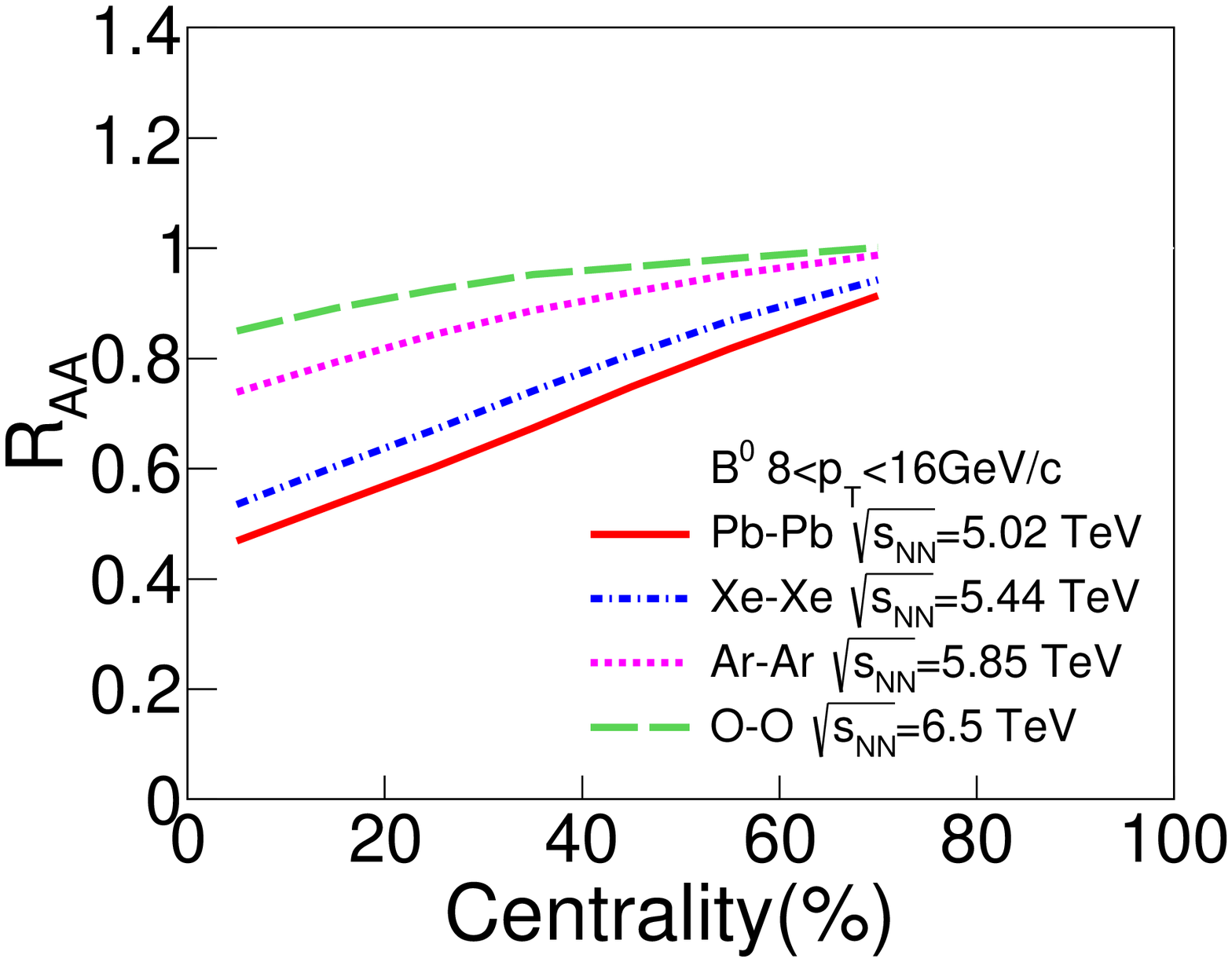}
    \hspace{-15pt}
    \includegraphics[clip=,width=0.25\textwidth]{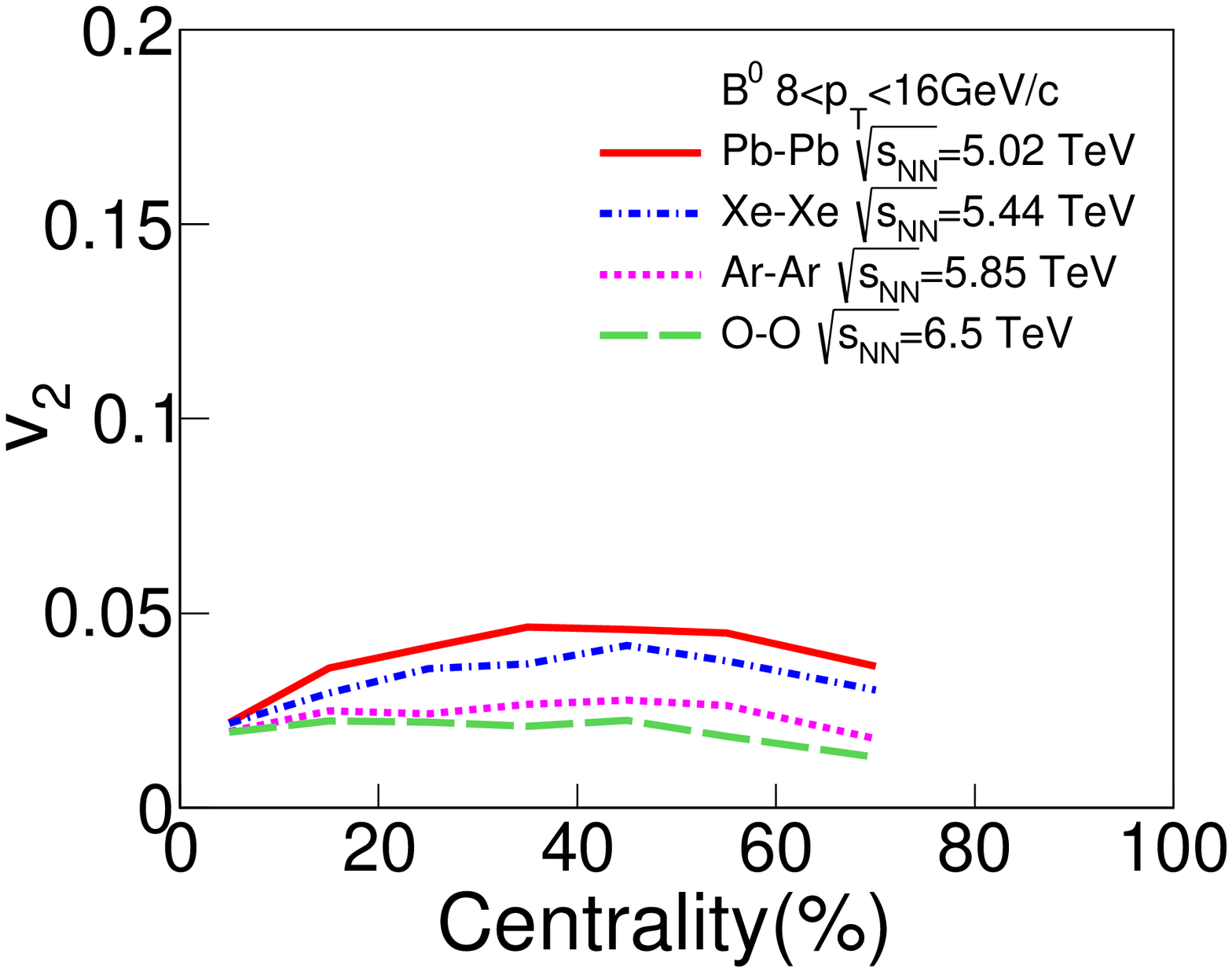}
    \caption{(Color online) Centrality dependence of $R_\mathrm{AA}$ (left) and $v_2$ (right) of $B$ mesons in different collision systems, upper for $5<p_\mathrm{T}<8$~GeV and lower for $8<p_\mathrm{T}<16$~GeV.}
    \label{fig:raa-v2-centrality-B0}
\end{figure}

To investigate how the heavy flavor observables rely on the geometry of the QGP, we present $R_\mathrm{AA}$ and $v_2$ as functions of centrality in Figs.~\ref{fig:raa-v2-centrality-D0} and~\ref{fig:raa-v2-centrality-B0} for $D$ and $B$ mesons respectively. Similar to the $N_\mathrm{part}$ dependence figures presented above, we show in each figure the $p_\mathrm{T}$-integrated observables within $5<p_\mathrm{T}<8$~GeV in the upper panel and $8<p_\mathrm{T}<16$~GeV in the lower panel. The left panels are for $R_\mathrm{AA}$ and the right for $v_2$. For a given collision system, we generally observe that the heavy flavor meson $R_\mathrm{AA}$ increases from central co peripheral collisions due to smaller heavy quark energy loss in more peripheral collisions, while $v_2$ first increases and then decreases due to the competing effects between parton energy loss and medium geometry. The only exception here is the large $D$ meson $v_2$ in central O-O collisions. This could be caused by the larger initial state fluctuations in smaller O nuclei, which generates large average eccentricity for the QGP fireballs produced in central O-O collisions.

Comparing different collision systems, one can clearly observe the hierarchies of both $R_\mathrm{AA}$ and $v_2$ of heavy flavor mesons. As discussed earlier, for a given centrality class (or medium eccentricity), a larger collision system has a higher $N_\mathrm{part}$, resulting in stronger energy loss of heavy quarks through the medium. This yields Pb-Pb $<$ Xe-Xe $<$ Ar-Ar $<$ O-O for the heavy meson $R_\mathrm{AA}$, and Pb-Pb $>$ Xe-Xe $>$ Ar-Ar $>$ O-O for their $v_2$. The exception of the $D$ meson $v_2$ in central O-O collisions is again caused by larger fluctuation effects in smaller collisions systems. Again, comparing Fig.~\ref{fig:raa-v2-centrality-D0} and~\ref{fig:raa-v2-centrality-B0}, we notice that $D$ mesons have much smaller $R_\mathrm{AA}$ and much larger $v_2$ than $B$ mesons due to the mass dependence of charm and bottom quark energy loss through the QGP.

\begin{figure}[tbp]
    \centering
    \includegraphics[clip=,width=0.25\textwidth]{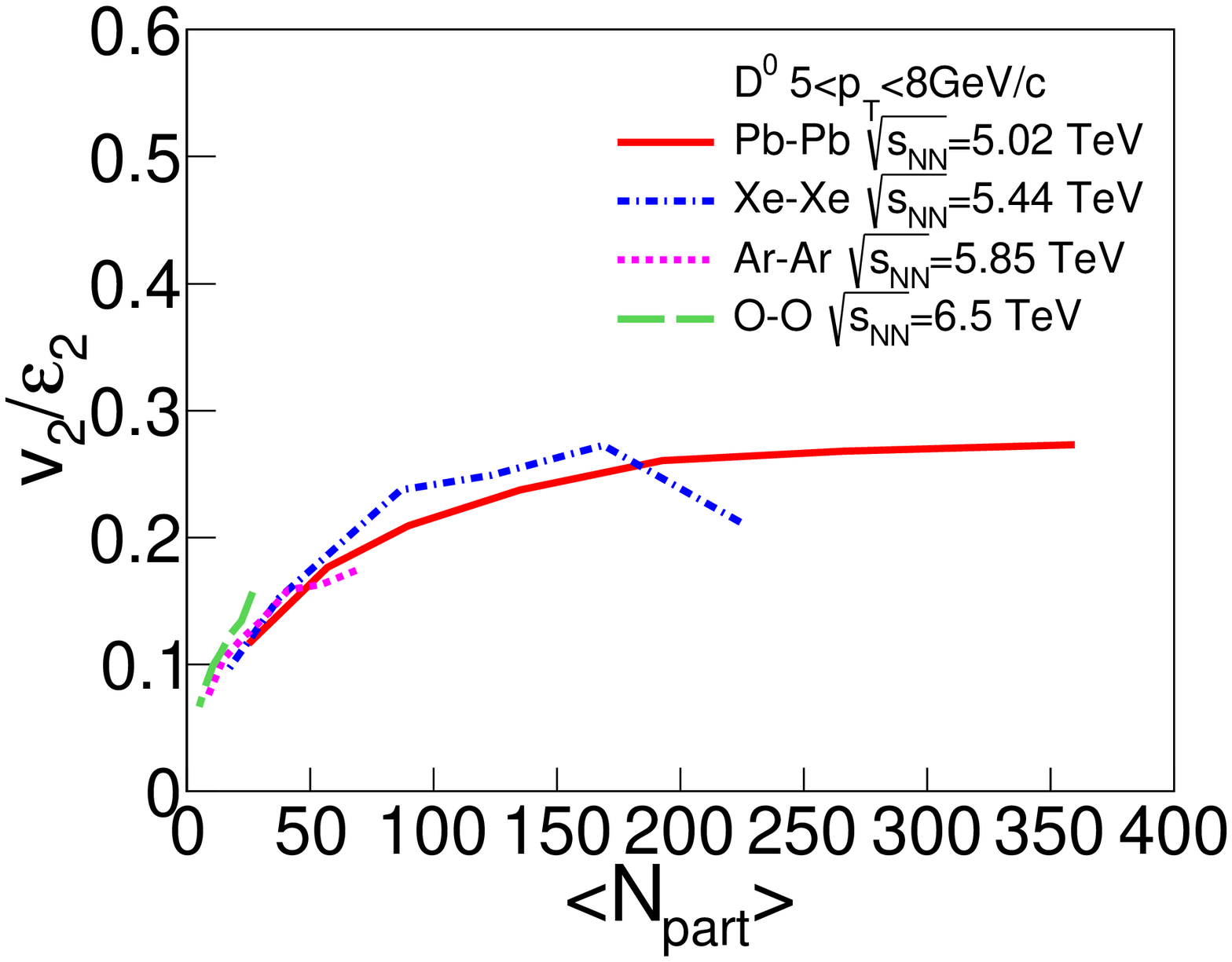}
    \hspace{-15pt}
    \includegraphics[clip=,width=0.25\textwidth]{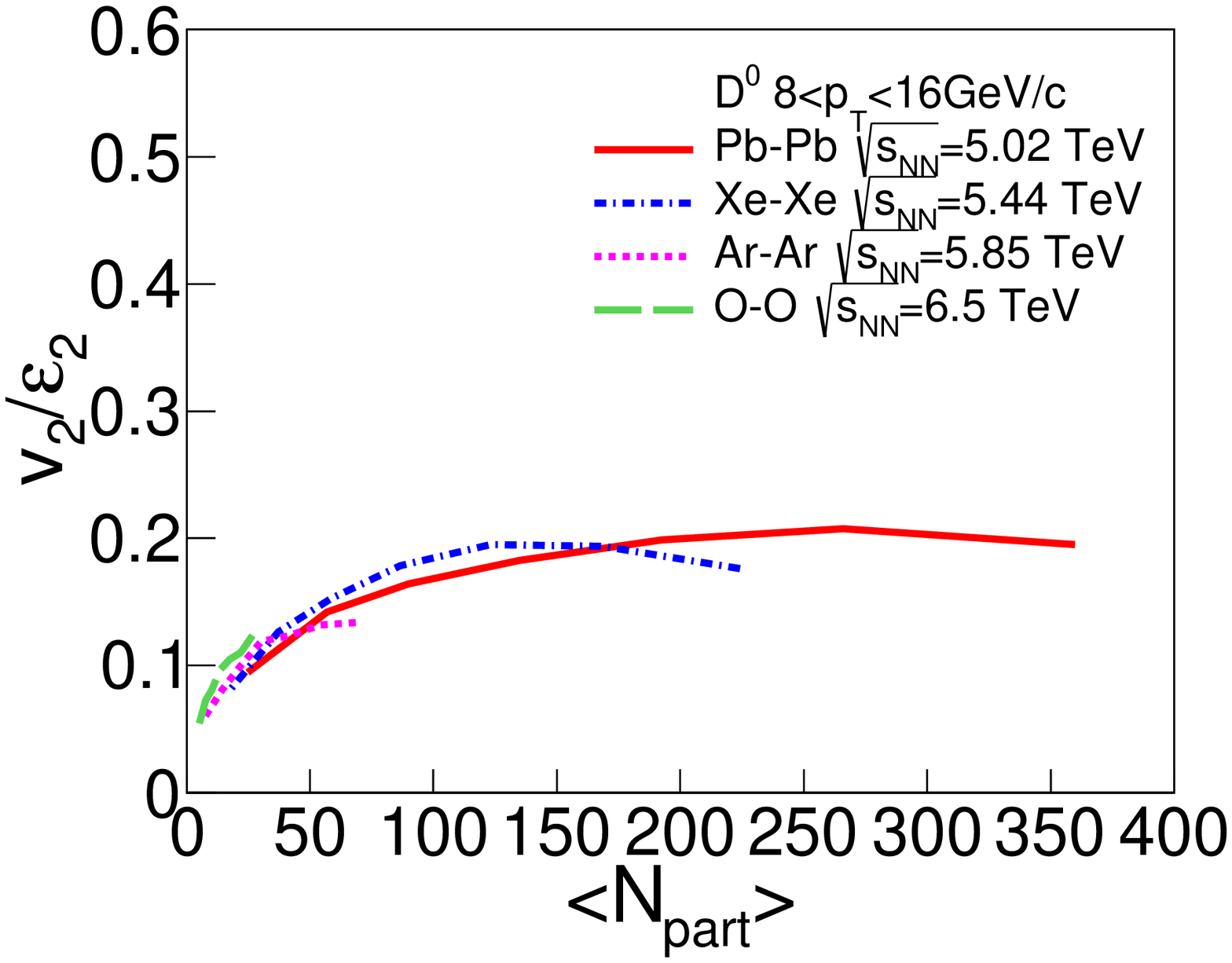}
    \includegraphics[clip=,width=0.25\textwidth]{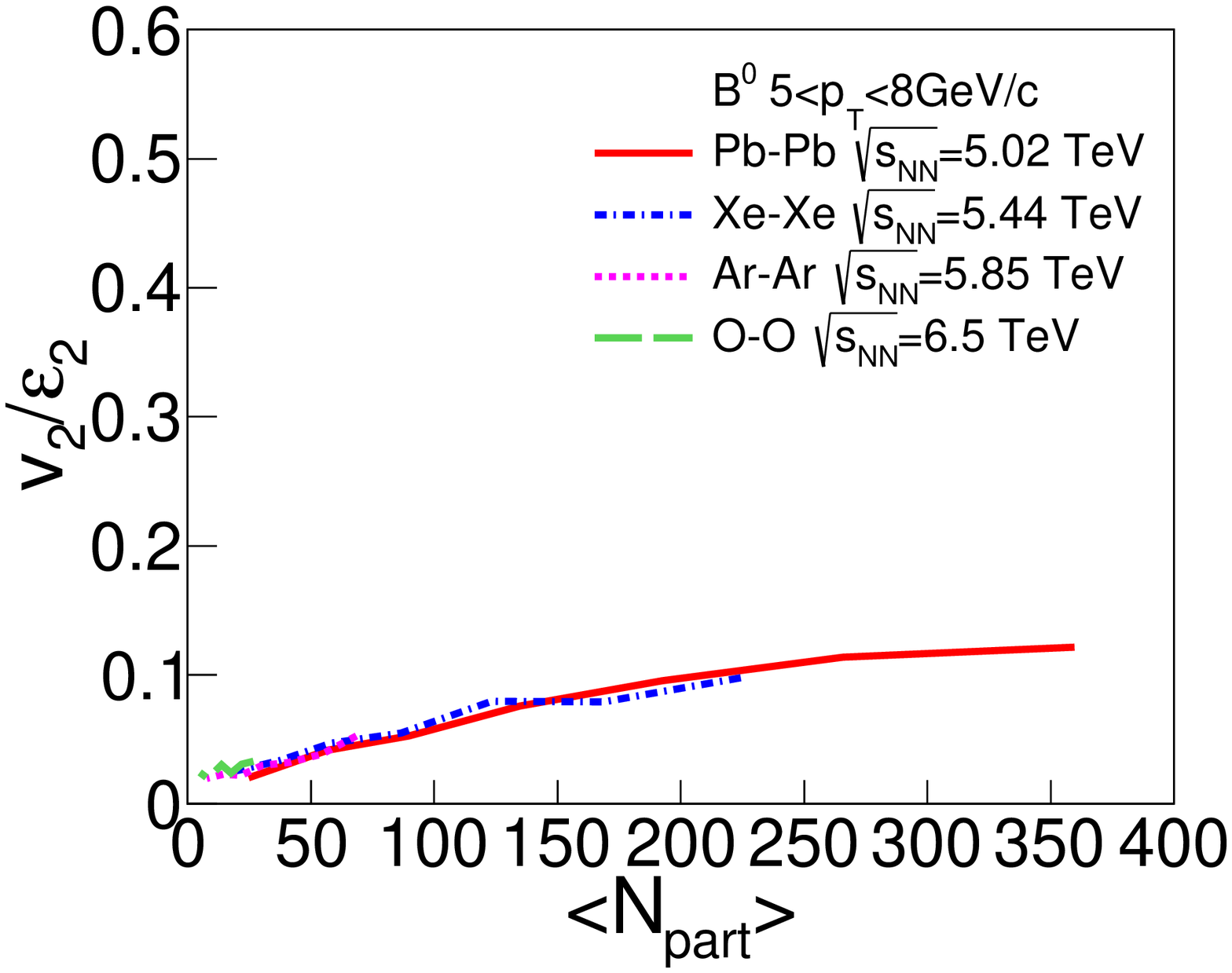}
    \hspace{-15pt}
    \includegraphics[clip=,width=0.25\textwidth]{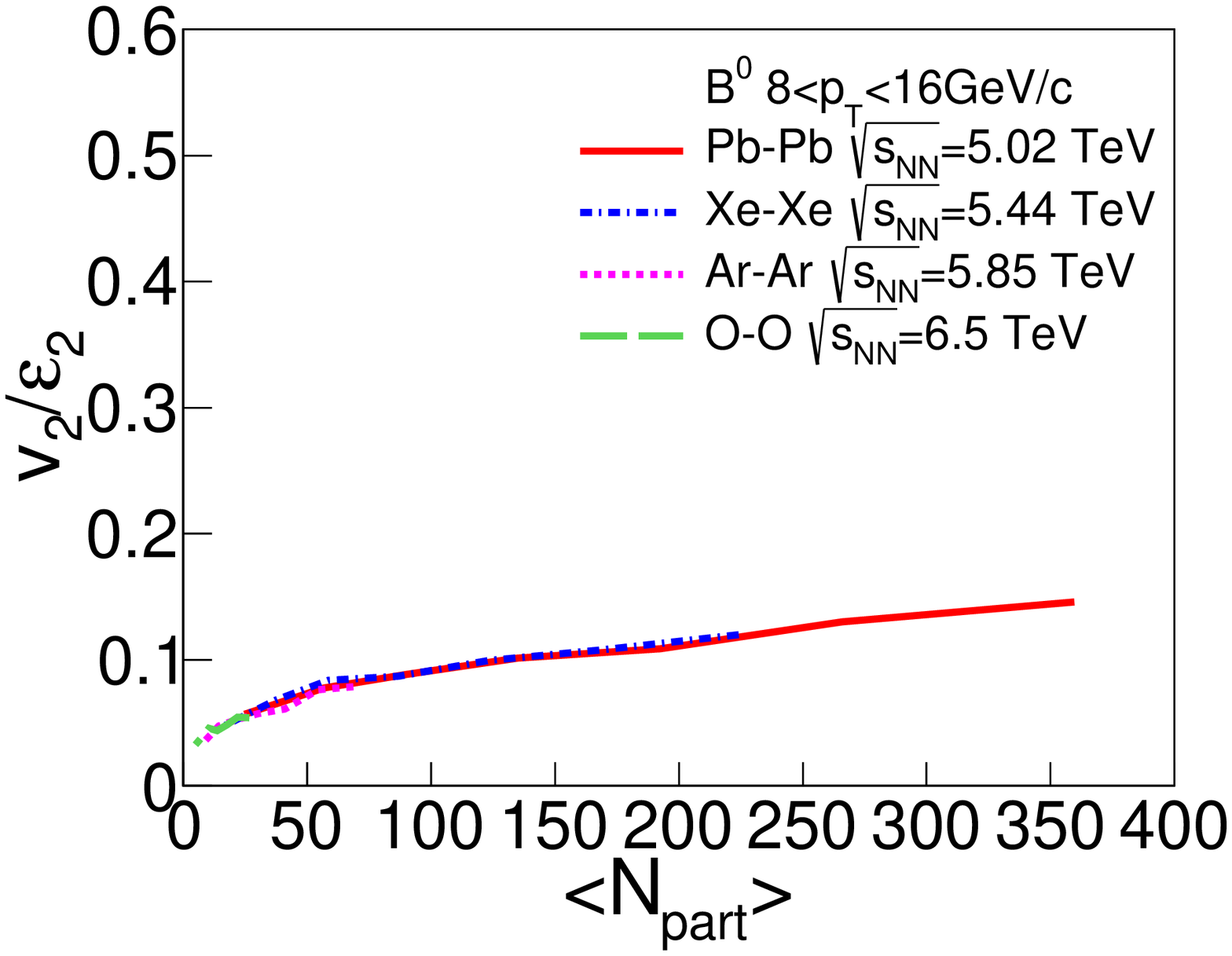}
    \caption{(Color online) Participant number dependence of the bulk $\varepsilon_2$ rescaled $v_2$ of $D$ mesons (upper) and $B$ mesons (lower) in different collision systems, left for $5<p_\mathrm{T}<8$~GeV and right for $8<p_\mathrm{T}<16$~GeV.}
    \label{fig:centralityv2-rescale}
\end{figure}

Since the elliptic flow coefficient $v_2$ of heavy flavor mesons depends on both the amount of heavy quark energy loss and the medium anisotropy, its scaling behavior between different collisions systems is hard to be displayed when plotting as a function of either $N_\mathrm{part}$ or centrality. {Interestingly, one may remove the medium anisotropy effect from the total contribution by rescaling the heavy meson $v_2$ with the bulk medium eccentricity $\varepsilon_2$. As shown in Fig.~\ref{fig:centralityv2-rescale}, the rescaled $v_2/\varepsilon_2$ is mainly determined by the amount of parton energy loss, thus scales with the system size or $N_\mathrm{part}$ between different collision systems. This behavior is very similar to heavy meson $R_\mathrm{AA}$. Note that although the amount of heavy quark energy loss is the main source of the heavy meson $v_2$ after removing the bulk geometry effect, the coupling of heavy quark motion to the QGP flow and the hadronization process can also affect the final state heavy meson $v_2$ and break the scaling behavior of $v_2/\varepsilon_2$ v.s. $N_\mathrm{part}$. Such breaking effect is more prominent for low energy heavy quarks and when the bulk radial flow effect is strong.}

\section{Summary}
\label{sec:summary}

Within our Langevin-hydrodynamics framework, we have performed a systematic study on the system size dependence of heavy quark energy loss in heavy-ion collisions at the LHC energies. The space-time evolution of the QGP produced in different collision systems is simulated using our (3+1)-dimensional CLVisc hydrodynamic model. The medium modification of the heavy quark energy-momentum is described by our modified Langevin equation that incorporates both elastic and inelastic scatterings of heavy quarks inside the QGP. By combining this Langevin model with the FONLL calculation for the initial heavy quark spectra and the fragmentation-coalescence model for hadronization, we have calculated the nuclear modification factor ($R_\mathrm{AA}$) and elliptic flow coefficient ($v_2$) of $D$ and $B$ mesons in various centrality regions of Pb-Pb, Xe-Xe, Ar-Ar and O-O collisions at the LHC. The transverse momentum, participant number and centrality dependences of the heavy meson $R_\mathrm{AA}$ and $v_2$ have been investigated in detail.

Our results show a clear system size dependence of the heavy meson $R_\mathrm{AA}$. For the same collision system, $R_\mathrm{AA}$ increases from central to peripheral collisions. For the same centrality class, $R_\mathrm{AA}$ increases (Pb-Pb $<$ Xe-Xe $<$ Ar-Ar $<$ O-O) as the size of colliding nuclei decreases. We have demonstrated a clear scaling of the heavy meson $R_\mathrm{AA}$ as a function of $N_\mathrm{part}$ between different collision systems, which indicates a direct correlation between the amount of jet energy loss and the size of the QGP profiles. On the other hand, the heavy meson $v_2$ simultaneously depends on the size and anisotropy of the QGP. For the same collision system, $v_2$ first increases and then decreases from central to peripheral collisions. For the same centrality class, $v_2$ generally increases as the size of colliding nuclei increases, except for the relatively large $v_2$ in central O-O collisions due to the strong initial-state fluctuations of the small O nucleus. {After eliminating the effects of different bulk medium anisotropy in different collision systems, the bulk-eccentricity-rescaled heavy meson elliptic flow ($v_2/\varepsilon_2$) is found to scale with $N_\mathrm{part}$. This reveals a direct correlation between $v_2/\varepsilon_2$ and the amount of heavy quark energy loss which depends on the overall size of QGP.} Moreover, the comparison between $D$ and $B$ mesons demonstrates a clear mass dependence of parton energy loss that yields smaller $R_\mathrm{AA}$ and larger $v_2$ of $D$ mesons than $B$ mesons for the same collisions system and the same centrality class.

The system size dependence of $D$ and $B$ meson observables discussed in this work provides a crucial bridge between large (Pb-Pb) and small (p-Pb) systems of relativistic nuclear collisions. Comparison between our numerical predictions here and future system-size-scan experiments on jet quenching is expected to help resolve several open questions in high-energy nuclear physics, such as the precise path-length dependence and mass dependence of parton energy loss, and the detailed correlation of collective flow coefficients between hard probes and soft hadrons. Interestingly, our calculation shows considerable amount of heavy quark energy loss even in the small O-O collisions, as suggested by the quenching effect on $R_\mathrm{AA}$ as well as the finite $v_2$ of both $D$ and $B$ mesons in central O-O collisions. This further implies that $R_\mathrm{pA}\sim 1$ in proton-nucleus collisions ~\cite{ALICE:2019fhe, Xu:2015iha} is mainly due to the small size of the nuclear medium in these even smaller collision systems. {We note that our earlier study~\cite{Xu:2015iha} suggests that the strong elliptic flow~\cite{CMS:2018loe} of $D$ mesons in p-Pb collisions cannot be explained by the final state parton energy loss effects.
In contrast, Refs.~\cite{Zhang:2019dth, Zhang:2020ayy} show that the initial state gluon saturation effect can explain well the observed elliptic flow of heavy mesons in p-Pb collisions.
Further investigations on both heavy and light flavor $R_\mathrm{AA}$ and $v_2$ in large and small systems, and their scaling behaviors, may help to identify the boundary where QGP disappears.}

\section*{Acknowledgments}

This work was supported by the Natural Science Foundation of China (NSFC) under Grants No. 11805082, No. 11775095, No. 11890710, No. 11890711 and No. 11935007, and Higher Educational Youth Innovation Science and Technology Program of Shandong Province (2019KJJ010).

\bibliographystyle{h-physrev5}
\bibliography{SCrefs}

\end{document}